# Spectroscopic approaches for studies of site-specific DNA base and backbone 'breathing' using exciton-coupled dimer-labeled DNA

Andrew H. Marcus,[1,2,*] Spiridoula Matsika,[3] Dylan Heussman,[1,2] Mohammed I. Sorour,[3] Jack Maurer,[1,2] Claire S. Albrecht,[4,5] Lulu Enkhbaatar,[1,2] Patrick Herbert,[1,2] Kurt A. Kistler[6] and Peter H. von Hippel[2]

[1.] Center for Optical, Molecular and Quantum Science, Department of Chemistry and Biochemistry, University of Oregon, Eugene, Oregon 97403

[2.] Institute of Molecular Biology, Department of Chemistry and Biochemistry, and University of Oregon, Eugene, Oregon 97403

[3.] Department of Chemistry, Temple University, Philadelphia, Pennsylvania 19122

[4.] Center for Optical, Molecular and Quantum Science, Department of Physics, University of Oregon, Eugene, Oregon 97403

[5.] Institute of Molecular Biology, Department of Physics, and University of Oregon, Eugene, Oregon 97403

[6.] Department of Chemistry, Brandywine Campus, The Pennsylvania State University, Media, Pennsylvania 19063

[*] Email: ahmarcus@uoregon.edu

**Abstract**

DNA regulation and repair processes require direct interactions between proteins and DNA at specific sites. Local fluctuations of the sugar-phosphate backbones and bases of DNA (a form of DNA 'breathing') play a central role in such processes. Here we review the development and application of novel spectroscopic methods and analyses – both at the ensemble and single-molecule levels – to study structural and dynamic properties of exciton-coupled cyanine and fluorescent nucleobase analogue dimer-labeled DNA constructs at key positions involved in protein-DNA complex assembly and function. The exciton-coupled dimer probes act as 'sensors' of the local conformations adopted by the sugar-phosphate backbones and bases immediately surrounding the dimer probes. These methods can be used to study the mechanisms of protein binding and function at these sites.

# Outline

# I. Biological significance of DNA 'breathing' and its connection to functional protein-DNA interactions and site-specific free energy surfaces

Biochemical processes central to gene regulation and expression are often described in terms of *local*, site-specific interactions between nucleic acids and proteins. For example, during DNA replication, proteins recognize and bind to key sites at and near single-stranded (ss) – double-stranded (ds) DNA junctions to establish the 'trombone'-shaped framework of the functional DNA replication-elongation complex (or replisome) [1]. Because the various sub-assemblies of the DNA replisome must copy the entire paternal genome, regardless of specific DNA base sequences, the binding affinities and interactions with the DNA scaffold must be controlled primarily by the local secondary structures of the bases, base-pairs, and sugar-phosphate backbones at the ss-dsDNA fork junctions at which replication proteins bind and function.

The local conformations of DNA near ss-dsDNA junctions undergo thermally activated fluctuations (termed DNA 'breathing') within an unknown distribution of macrostates to permit the proper binding of replication proteins. These macrostates are perturbed relative to the canonical Watson-Crick (W-C) duplex structure [2], exhibiting one or more of the following characteristics: *i*) flanking bases may be 'twisted' relative to their neighbors and thus somewhat more or less stacked; *ii*) inter-base hydrogen bonds may be disrupted; and/or *iii*) flanking bases may be fully unstacked or un-hydrogen-bonded, thus exposing the base sequences to the external environment [3, 4]. Because non-canonical conformations are unstable relative to the W-C structure, they occur transiently at trace concentrations. Nevertheless, such non-canonical conformations likely act as rate-limiting intermediates for the initial chemical steps of recognition, assembly and function of the protein-DNA complexes involved in core biological processes [5].

To understand the mechanisms by which multicomponent complexes such as the DNA replisome self-assembles and functions, it is important to identify key features of the underlying macromolecular interactions. In Fig. 1 is shown a hypothetical free energy surface (FES) that depicts a possible cooperative assembly mechanism of the T4 bacteriophage helicase-primase (primosome)-DNA fork complex [6, 7]. The primosome is a sub-assembly of the T4 replisome, which is a useful model system to understand the replication complexes of higher organisms [8]. The FES describes variations in the stability of protein and nucleic acid components bound at specific interaction sites and as a function of the relevant chemical reaction coordinates (labeled $Q_1$, $Q_2$, ...). Points on the FES corresponding to free energy minima and maxima define the local conformations of stable macrostates and unstable transition states, respectively, which lie along the functional assembly pathway of the protein-DNA complex. 'Basins' that surround each local minimum of the FES are associated with the distributions of quasi-degenerate states, which define the conformational disorder of the macrostate.

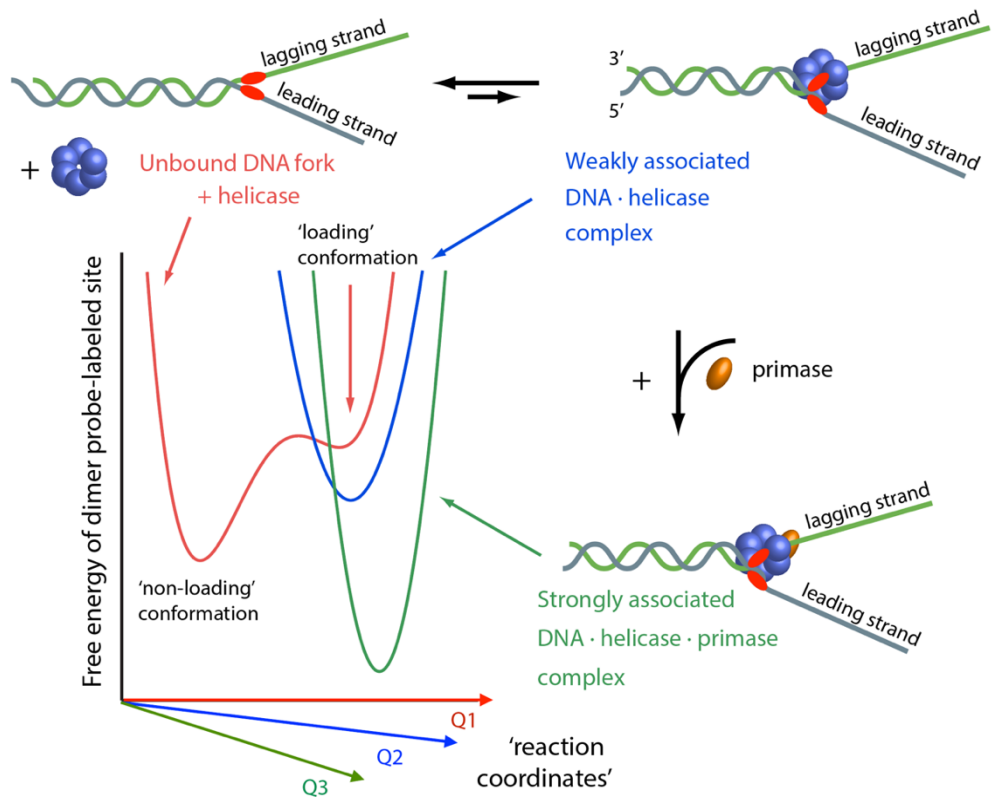

**Figure 1.** Hypothetical free energy surface (FES) for the assembly steps of the T4 helicase-primase (primosome)-DNA fork sub-assembly as monitored by exciton-coupled (iCy3)$_2$ dimer backbone probes (indicated by red ovals). Successive reaction steps are indicated by red, blue and green 2D cross sections of the multidimensional FES. Figure adapted from Ref. [2].

The FES shown in Fig. 1 emphasizes the role of DNA 'breathing' at a specific site near the ss-dsDNA fork junction where the helicase and primase must preferentially bind and assemble in a defined sequence of events. In the absence of protein, the DNA bases and sugar-phosphate backbones at a specific position near the ss-dsDNA fork junction may fluctuate between a stable conformation that does not favor functional helicase loading, and an unstable conformation that does favor such loading [2]. The association of a weakly bound helicase with the subsequent formation of a stable and fully functional primosome-DNA fork complex is represented as a series of successive 'free-energy-surface-crossing' events.

Many of the molecular details currently known about site-specific protein-DNA interactions – and how they participate in functional assembly mechanisms as depicted in Fig. 1 – has been inferred by applying powerful structural methods such as nuclear magnetic resonance (NMR), X-ray crystallography, and cryo-electron microscopy [9-13]. However, such methods are not well suited to study DNA 'breathing' of protein-DNA complexes *in situ* at 'physiological' concentrations and solution conditions that simulate the intracellular environment.[1] In the remainder of this chapter, we review novel optical spectroscopic approaches that can be applied to study the dynamics of protein-DNA interactions under biologically relevant conditions in which such complexes function.

**II. Exciton-coupled probe spectroscopy for studies of *in-situ* site-specific DNA 'breathing'**

An approach that is well suited for *in-situ* studies of protein-DNA complexes at 'physiological' concentrations is to perform spectroscopic experiments on a DNA molecule that has been site-specifically labeled with fluorescent optical probes [15-18]. For example, the fluorescent cyanine dyes Cy3 and Cy5 can be chemically attached to a nucleic acid base or to the distal end of a nucleic acid chain using a chemical linker. Such cyanine dye-labeled DNA constructs are useful – due largely to their relatively high absorption cross-sections and fluorescence quantum yields – for fluorescence microscopy, gene sequencing technologies, and other bio-analytical applications. Spectroscopic measurements based on Förster resonance energy transfer (FRET) of a DNA strand labeled with a Cy3 'donor' and a Cy5 'acceptor' can monitor global conformation changes on the length scale of a few nanometers [18-22]. Alternatively, pairs of Cy3 dye molecules can be attached 'internally' (termed 'iCy3') and rigidly at defined opposing positions within the sugar-phosphate backbones of complementary DNA strands to create an exciton-coupled $(iCy3)_2$ dimer-probe-labeled DNA duplex or ss-dsDNA construct. Such exciton-coupled $(iCy3)_2$ dimer-labeled DNA constructs can be used to 'sense' the local environment of the DNA bases and backbones within the immediate vicinity of the dimer probe.

---

[1] The 'physiological' salt concentrations referred to in this chapter are typically [NaCl] = 100 mM and [MgCl$_2$] = 6 mM for *in vitro* studies. These are not the actual intra-cellular salt concentrations, which fluctuate according to changing environmental conditions. Rather these values correspond to a set of salt concentrations that result in the same protein-DNA binding constants in *in vitro* model systems as those measured in living *E. coli* cells. 14. Y. Kao-Huang, A.R., A. P. Butler, P. O'Connor, D. W. Noble, P. H. von Hippel, *Nonspecific DNA binding of genome-regulating proteins as a biological control mechanism: Measurement of DNA-bound Escherichia coli lac repressor in vivo.* Proc. Nat. Acad. Sci., 1977. **74**: p. 4228-4232.

In Fig. 2*A* we show how iCy3 is attached internally within the framework of a DNA strand using phosphoramidite chemistry. The iCy3 label acts as a relatively stiff molecular bridge between DNA bases and as part of the sugar-phosphate backbones connecting adjacent nucleotide residues. The electric dipole transition moment (EDTM, shown as a double-headed green arrow) lies along the longitudinal direction of the π-conjugated trimethine group. By annealing two complimentary single strands of DNA with opposite iCy3 labeling positions, an exciton-coupled (iCy3)$_2$ dimer-labeled DNA backbone probe can be formed at a predetermined position within a DNA fork construct (see Fig. 2*B*). As discussed in the following sections, the absorption and circular dichroism (CD) spectra of an (iCy3)$_2$ dimer-labeled DNA construct are highly sensitive to the conformation of the dimer probe, which in turn depends sensitively on both the labeling position within the DNA construct and on environmental conditions (e.g., salt concentration, temperature, etc.) that affect the stabilities of the local DNA bases and sugar-phosphate backbones immediately adjacent to the probe [23-25].

Another probe labeling strategy that we discuss in later sections uses pairs of fluorescent DNA base analogues, which can be substituted for native DNA bases as a flanking dimer within the native DNA duplex at sites close to a ss-dsDNA junction [26-28]. In Fig. 2*C* and 2*D*, respectively, are shown molecular structures of the fluorescent base analogues 6-methyl isoxanthopterin (6-MI, an analogue of guanine) and 2-amino purine (2-AP, an analogue of adenine) [29-35]. Like their canonical nucleobase counterparts, 6-MI and 2-AP can form W-C base-pairs with their complementary cytosine and thymine bases, respectively. An example structure of a (6-MI)$_2$ dimer-substituted ss-dsDNA fork construct is shown in Fig. 2*E*. 6-MI and 2-AP absorb light at lower energy than the absorption of native nucleic acids (i.e., at wavelengths longer than ~300 nm, where the native bases are transparent) and they exhibit significant fluorescence quantum yields. Example absorption and fluorescence spectra of a 6-MI-substituted ss-dsDNA construct is shown in Fig. 2*F*. Two closely spaced fluorescent base analogues – incorporated as flanking bases within the same strand – can form an exciton-coupled dimer probe. Like the (iCy3)$_2$ dimer backbone probes discussed above, local conformational changes of the (6-MI)$_2$ dimer probe can be monitored using fluorescence, absorption and CD spectroscopy, as discussed in further detail below [26, 27, 35].

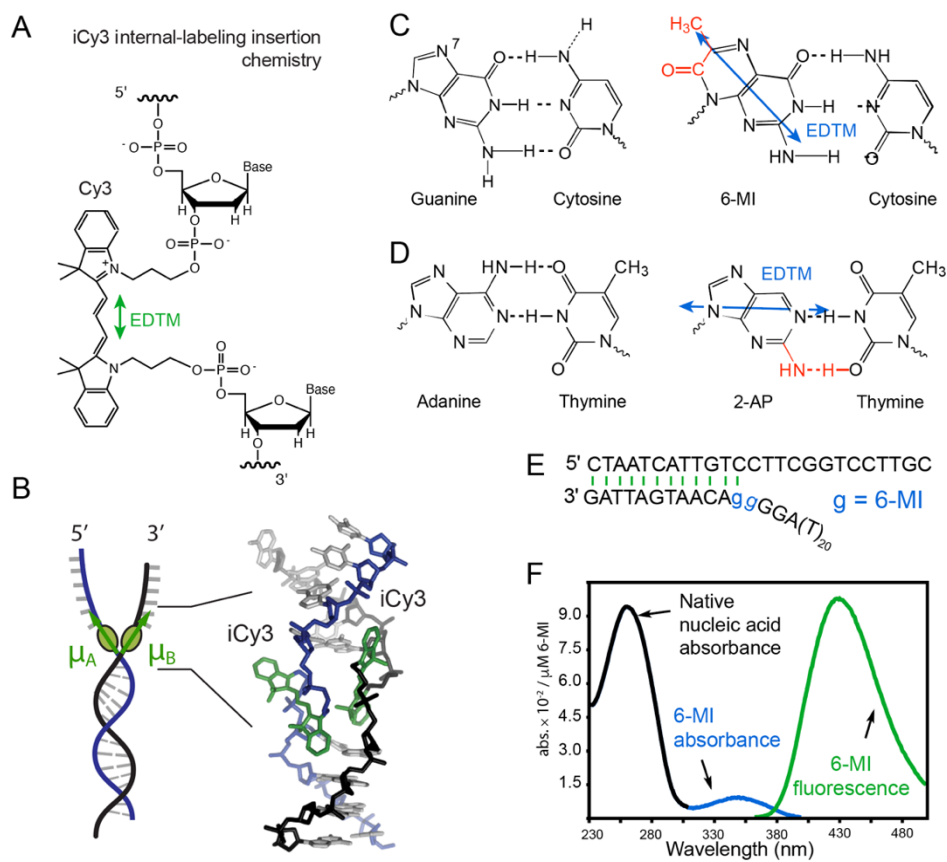

**Figure 2.** (*A*) Chemical structure of the internally labeled iCy3 chromophore, indicating the direction of its lowest energy electric dipole transition moment (EDTM) and its internal-labeling insertion chemistry. (*B*) Example of a DNA fork construct with an exciton-coupled (iCy3)$_2$ dimer probe labeling a specific position relative to the ss-dsDNA junction. (*C*) Structure of the G:C (left) versus the 6-MI:C (right) Watson-Crick base pair. (*D*) Structure of the A:T (left) versus the 2-AP:T base pair (right). The difference between the natural bases and the analogues are shown in red. The directions of the lowest energy EDTMs are shown in blue. (*E*) An example of a DNA fork construct labeled with two adjacent 6-MI nucleotide analogues (indicated by gg in blue) positioned at specific positions relative to the fork junction. (*F*) Absorption and fluorescence spectra of 6-MI substituted DNA.

## III. Exciton-coupled (iCy3)$_2$ dimer-labeled ss-dsDNA constructs as probes of local DNA base and backbone conformations

An annealed (iCy3)$_2$ dimer-labeled DNA fork construct contains both double-stranded and single-stranded DNA regions, and the dimer probe can be selectively inserted relative to the ss-dsDNA fork junction. In Fig. 3*A* is shown the nomenclature that we use to specify the probe-labeling position relative to the ss-dsDNA fork junction with progressively increasing positive integers indicating positions toward the double-stranded region and negative integers indicating positions toward the single-stranded regions. An iCy3 monomer-labeled ss-dsDNA construct can also be prepared by including a thymine (T) base in the complementary strand at the position directly opposite to the probe chromophore. The addition of the T base partially compensates for the misalignment of complementary base pairs introduced by the presence of the iCy3 monomer probe within one strand.

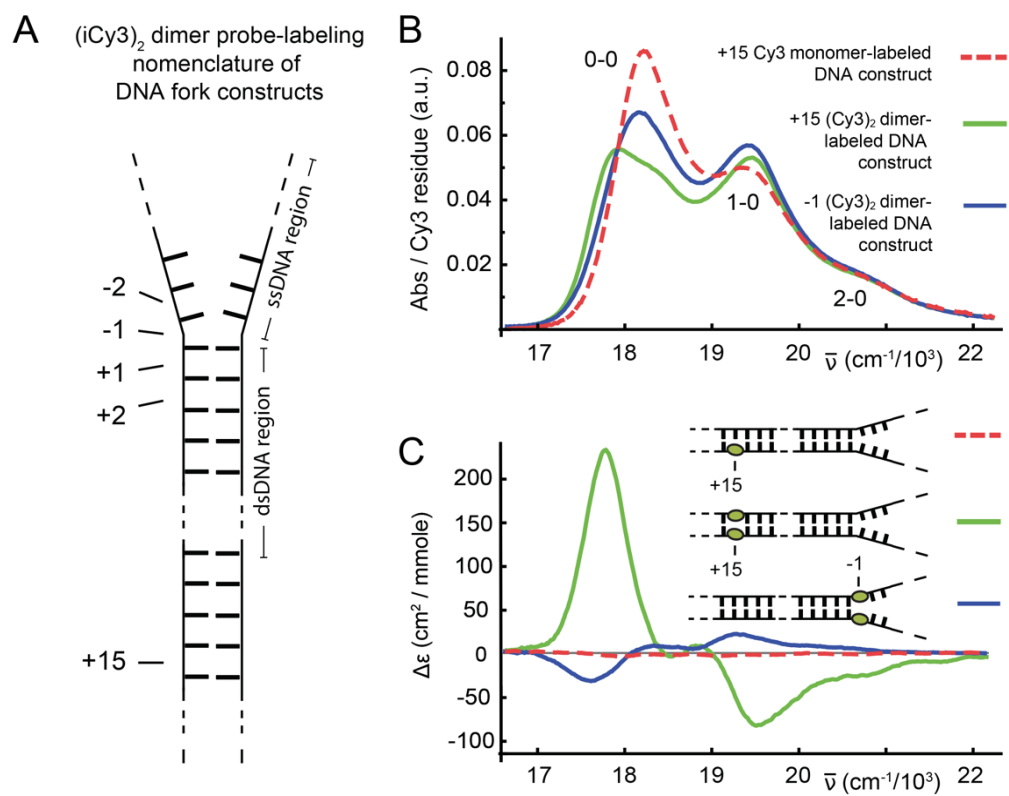

**Figure 3.** (*A*) The insertion site position of the (iCy3)$_2$ dimer probes is indicated relative to the ss-dsDNA fork junction using positive integers in the direction toward the double-stranded region and negative integers in the direction toward the single-stranded region. (*B*) Room temperature (23°C) absorption and (*C*) CD spectra for iCy3 monomer-labeled +15 duplex (dashed red curve), (iCy3)$_2$ dimer-labeled +15 duplex (green), and (iCy3)$_2$ dimer-labeled -1 fork (blue) DNA constructs. Measurements were performed at 23°C using 100 mM NaCl and 6 mM MgCl$_2$, and 10 mM Tris at pH 8.0. Figure adapted from Ref. [23].

In Figs. 3*B* and 3*C* are shown experimental (room temperature, 23°C) absorption and CD spectra of iCy3 monomer- and (iCy3)$_2$ dimer-labeled DNA constructs with probes inserted deep within the duplex region (+15 position), and an (iCy3)$_2$ dimer-labeled DNA construct with probes inserted near the ss-dsDNA fork junction (-1 position). The linear absorption spectrum of the iCy3 monomer (red dashed curve) exhibits a pronounced electronic-vibrational (vibronic) progression with progressively increasing energy features (labeled 0-0, 1-0 and 2-0), while its CD spectrum is very weak. The absorption and CD of the iCy3 monomer are largely insensitive to factors that affect the local environment of the chromophore, such as temperature, solution salt concentrations, and probe-labeling position relative to the ss-dsDNA fork junction [23, 24, 36, 37]. In contrast, the absorption and CD spectra of the (iCy3)$_2$ dimer-labeled DNA constructs are highly sensitive to probe labeling position and environmental factors that affect the local probe conformation. The vibronic progression is still present in the spectra of both the +15 and -1 (iCy3)$_2$ dimer-labeled ss-dsDNA constructs (solid green and blue curves, respectively). However, the individual vibronic features of the (iCy3)$_2$ dimers are broadened relative to those of the iCy3 monomer, and the ratios of the 0-0 to 1-0 vibronic peak intensities are smaller compared to the monomer $[I_{\text{mon}}^{(0-0)}/I_{\text{mon}}^{(1-0)} = 1.60]$ [36]. The CD of both the +15 and -1 (iCy3)$_2$ dimer-labeled ss-dsDNA constructs exhibit a progression of bisignate lineshapes (*i.e.*, a change of sign within a given vibronic band), which is a signature of vibronic excitons in a chiral dimer [38]. Most notably, the CD spectra of the +15 and -1 (iCy3)$_2$ dimer-labeled ss-dsDNA constructs have opposite signs, indicating that the two systems have opposite chiral symmetries.

## III.A Characterization of mean local conformational parameters from absorption and CD spectra of exciton-coupled (iCy3)$_2$ dimer-labeled ss-dsDNA fork constructs

The Cy3 chromophore consists of a conjugated trimethine bridge, which cojoins two indole-like groups, as shown in Fig. 2*A*. The trimethine bridge of the Cy3 monomer exists in an *all-trans* configuration, and the lowest energy electronic transition is a $\pi \to \pi^*$ transition [39]. Like many $\pi$-conjugated molecules, Cy3 exhibits numerous Franck-Condon active modes, which range in energy from tens-to-several hundreds of wavenumbers [40]. Nevertheless, the homogeneous spectral lineshape of the Cy3 monomer in solution is broadened due to rapid electronic dephasing and thermal occupation of low energy vibrational levels in the electronic ground state. The lowest energy electronic transition from ground state $|g\rangle$ to excited state $|e\rangle$ is predominantly coupled to one low-frequency (~30 cm$^{-1}$) symmetric bending vibration along the trimethine bridge, which in turn couples anharmonically to a cluster of relatively high-frequency modes in the vicinity of ~1,300 cm$^{-1}$ [37], as described in further detail in Sect. III.B.

The spectrum of the Cy3 monomer in solution – and that of the iCy3 monomer inserted within the DNA sugar-phosphate backbone – can be simulated using a relatively simple quantum mechanical model Hamiltonian, $\widehat{H}_M$, where the $|g\rangle \to |e\rangle$ transition with energy $\varepsilon_{eg}$ is coupled to a single 'effective' harmonic mode with energy $\hbar\omega_{\text{vib}}$, as shown in Figs. 4*A* and 4*B* [23-25, 36]. The monomer Hamiltonian is written:

$$\widehat{H}_M = \varepsilon_{eg}|e\rangle\langle e| + \hbar\omega_{\text{vib}}\hat{b}_M^\dagger\hat{b}_M + \hbar\omega_{\text{vib}}\{\lambda(\hat{b}_M^\dagger + \hat{b}_M) + \lambda^2\}|e\rangle\langle e| \qquad (1)$$

where $\hat{b}_M^\dagger$ and $\hat{b}_M$ are, respectively, the operators for creating and annihilating a vibrational excitation in the ground and excited electronic potential energy surfaces. These operators obey the commutation relation $[\hat{b}_{M'}, \hat{b}_M^\dagger] = \delta_{M',M}$, where $\delta_{M',M}$ is the Kronecker delta function and the monomer labels $M', M \in \{A, B\}$.

The strength of the vibronic coupling is characterized by Franck-Condon (FC) factors, $|\langle n_e|0\rangle|^2 = e^{-\lambda^2}\lambda^{2n_e}/n_e!$, where $|0\rangle = |n_g = 0\rangle|g\rangle$ is the zero vibrational level of the ground electronic state and $|n_e\rangle$ is the vibrational quantum number of the electronically excited state with vibrational quantum number $n_e$. The FC factors depend on the Huang-Rhys parameter, $\lambda^2 = d^2\omega_{\text{vib}}m/2\hbar$ where $d$ is the displacement of the electronically excited vibrational potential energy surface relative to the ground state (see Fig. 4*A*) surface and $m$ is the reduced mass of the effective mode [25]. The Huang-Rhys parameter also determines the vibrational relaxation energy, $\lambda^2\hbar\omega_{\text{vib}}$, which is the amount of optical transition energy that is converted into heat.

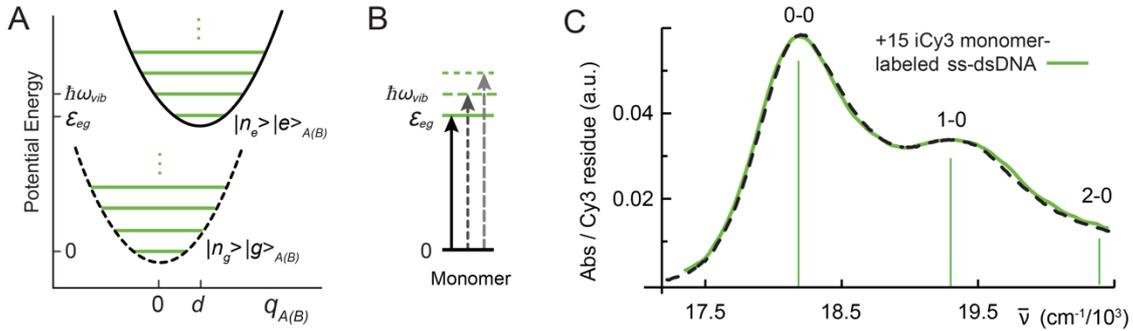

**Figure 4.** (***A***) Electronic-vibrational (vibronic) potential energy diagram for the iCy3 monomer [labeled $A(B)$] ground and excited electronic state levels [labeled $|g\rangle_{A(B)}$ and $|e\rangle_{A(B)}$, respectively], which are assumed to be coupled to a single 'effective' high-frequency vibrational mode with ground and excited vibrational states [labeled $|n_g\rangle_{A(B)}$ and $|n_e\rangle_{A(B)}$, respectively]. (***B***) Simplified energy level diagram of the iCy3 monomer. (***C***) Experimental and simulated absorption spectrum of +15 iCy3 monomer-labeled ss-dsDNA construct. Individual vibronic sub-bands are labeled 0-0, 1-0 and 2-0, as shown, indicating the vibrational levels in the ground and excited electronic states. Figure adapted from Ref. [41].

In Fig. 4*C* are shown the experimental (green curve) and simulated (dashed black) absorption spectrum of the +15 iCy3 monomer-labeled ss-dsDNA construct at room temperature (23°C). The simulated monomer spectrum accounts for the presence of both homogeneous and inhomogeneous line broadening. The homogeneous spectrum, $\sigma_H^M(\varepsilon)$, is the sum of homogeneous lineshapes associated with the individual vibronic transitions [24]

$$\sigma_H^M(\varepsilon) = |\mu_{eg}|^2 \sum_{n_e=0}^{\infty} |\langle n_e|0\rangle|^2 L_H(\varepsilon - \varepsilon_{eg} - n_e\hbar\omega_{\text{vib}}) \qquad (2)$$

Individual homogeneous lineshapes are modeled as Lorentzian functions, $L_H(\varepsilon) = \frac{1}{2}\Gamma_H / \left[\varepsilon^2 + \left(\frac{1}{2}\Gamma_H\right)^2\right]$, with full-width-at-half-maximum (FWHM) equal to $\Gamma_H$. The magnitude of the homogeneous linewidth is due to coupling between the optical transition and the phonon bath, which rapidly modulates the monomer transition energy. The value of $\Gamma_H = 186$ cm$^{-1}$ for the iCy3 monomer was determined using two-dimensional fluorescence spectroscopy (2DFS) [24, 36]. The magnitude of the iCy3 monomer EDTM, $\mu_{eg} = \sim 12.8$ D, was determined by integrating numerically the experimental absorption lineshape [36].

Although each monomer is chemically identical, variations in the local environment can lead to the presence of static inhomogeneity of the transition energy, $\varepsilon_{eg}$ [23, 24]. We modeled static inhomogeneity using a Gaussian distribution, $G_{I,\text{mon}}(\varepsilon_{eg}) = \exp\left[-(\varepsilon_{eg} - \bar{\varepsilon}_{eg})^2/2\sigma_{I,\text{mon}}^2\right]$, with mean transition energy $\bar{\varepsilon}_{eg}$ and spectral inhomogeneity standard deviation $\sigma_{I,\text{mon}}$. The simulated absorption spectrum is thus a convolution integral of homogeneous and inhomogeneous line shape functions [24]:

$$\sigma_I^M(\varepsilon) = \int_{-\infty}^{\infty} \sigma_H^M(\varepsilon - \varepsilon') G_{I,\text{mon}}(\varepsilon') d\varepsilon' \qquad (3)$$

As discussed above, the simulated absorption spectrum of the iCy3 monomer depends on four independent model parameters: (*i*) the monomer electronic transition energy, $\varepsilon_{eg}$; (*ii*) the Huang-Rhys electronic-vibrational coupling parameter, $\lambda^2$; (*iii*) the effective vibrational energy, $\hbar\omega_{\text{vib}}$; and (*iv*) the spectral inhomogeneity parameter $\sigma_{I,\text{mon}}$. These parameters were determined by fitting the above model to experimental spectra of the +15 iCy3 monomer-labeled DNA construct from which we determined the following values: $\varepsilon_{eg} = \sim 18,277$ cm$^{-1}$, $\lambda^2 = \sim 0.56$, $\hbar\omega_{\text{vib}} = \sim 1,109$ cm$^{-1}$, and $\sigma_{I,\text{mon}} = \sim 347$ cm$^{-1}$ [36]. We note that the vibrational relaxation energy corresponding to these values is $\lambda^2 \hbar\omega_{\text{vib}} = \sim 605$ cm$^{-1}$. As mentioned above, these values for iCy3 monomer-labeled ss-dsDNA constructs are relatively insensitive to labeling position and other environmental factors.

For the (iCy3)$_2$ dimer-labeled ss-dsDNA constructs, the closely-spaced iCy3 monomers can couple through a transition density interaction, $J$ [24, 25, 36, 42, 43]. The value of $J$ and the resulting spectroscopic properties of the (iCy3)$_2$ dimer-labeled DNA construct depend sensitively on the relative orientation and spacing between the iCy3 monomers on the length scale of a few Angstroms (see, for example, Figs. 3*B* and 3*C*), which in turn depends on the local stabilities and dynamics of the sugar-phosphate backbones and bases immediately adjacent to the probes.

The Holstein-Frenkel (H-F) Hamiltonian of the coupled dimer is written [44]

$$\hat{H}_A + \hat{H}_B + J[|eg\rangle\langle ge| + |ge\rangle\langle eg|] \tag{4}$$

where $|eg\rangle$ is the product state in which monomer *A* is electronically excited and monomer *B* is unexcited. For non-zero electrostatic coupling $J$, the absorption spectrum of the (iCy3)$_2$ dimer has a shape roughly like that of the monomer. However, the electrostatic interaction leads to each of the vibronic features (i.e., 0-0, 0-1, ...) being split and additionally broadened into symmetric (+) and anti-symmetric (−) sub-bands, as illustrated in Fig. 5*A*. The relative peak intensities of the absorption spectrum depend on – in addition to the Franck-Condon factors that affect the vibronic features of the iCy3 monomer – the local conformation of the dimer, which determines the magnitudes of the symmetric and anti-symmetric transition dipole moments $\boldsymbol{\mu}_\pm = \frac{\mu_{eg}}{\sqrt{2}}[\hat{\boldsymbol{d}}_A \pm \hat{\boldsymbol{d}}_B]$, where $\hat{\boldsymbol{d}}_A$ and $\hat{\boldsymbol{d}}_B$ are unit vectors that specify the monomer EDTM directions (see Figs. 5*B* and 5*C*). The symmetric and anti-symmetric excitons each consist of a manifold of delocalized electronically-coupled vibronic states, which are superpositions of electronic-vibrational products of the *A* and *B* monomer states given by $\left|e_\pm^{(\alpha)}\right\rangle = \sum_{n_e,n_g} c_\pm^{(\alpha)} \left[u_{n_e,n_g}|eg\rangle \pm u_{n_g,n_e}|ge\rangle\right]$ [45, 46]. The coefficients, $c_\pm^{(\alpha)}$, depend on the vibrational coordinates of the monomers, $u_{n_{e(g)},n_{g(e)}} = |n_{e(g)}n_{g(e)}\rangle$ is the vibrational product state, $n_{e(g)}$ is the vibrational quantum number in the electronic excited (ground) state, and $\alpha = (0, 1, …)$ is a state index in order of increasing energy.

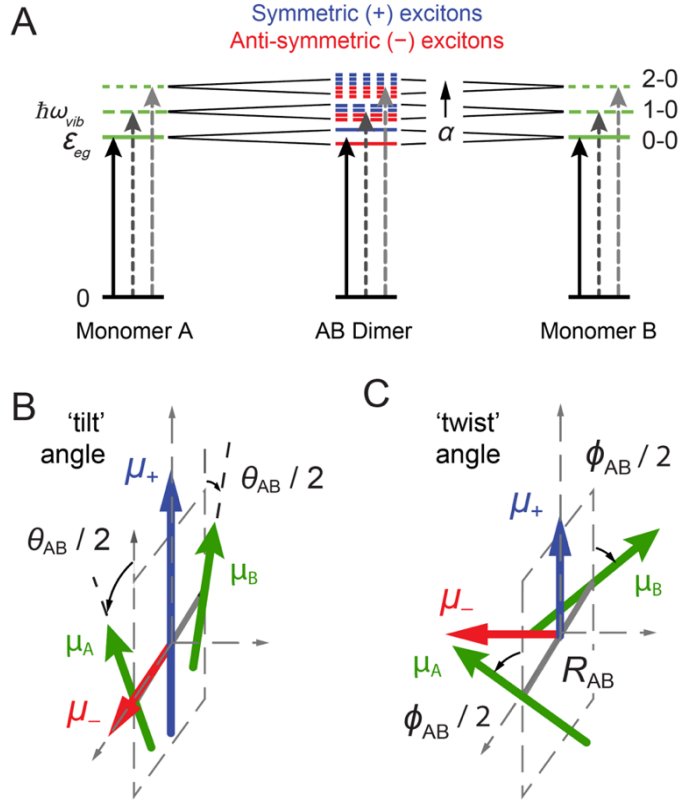

**Figure 5.** (*A*) Energy level diagrams of separated iCy3 monomers [labeled *A(B)*] the coupled (iCy3)$_2$ dimer. The electronic coupling (*J*) between closely spaced monomers of the (iCy3)$_2$ dimer results in delocalized symmetric (+, blue) and anti-symmetric (−, red) excitons. The dimer structural parameters are (*B*) the 'tilt' angle, $\theta_{AB}$, (*C*) the 'twist' angle, $\phi_{AB}$, and the interchromophore separation, $R_{AB}$. The electric dipole transition moments of the monomers [$\mu_{A(B)}$] are indicated by green arrows, and those of the symmetric ($\mu_+$) and anti-symmetric ($\mu_-$) excitons are indicated by blue and red arrows, respectively. Figure adapted from Ref. [41].

The (iCy3)$_2$ dimer homogeneous absorption spectrum is the sum of symmetric (+) and anti-symmetric (−) excitons:

$$\sigma_H^D(\varepsilon) = \sigma_+(\varepsilon) + \sigma_-(\varepsilon) \qquad (5)$$

where $\sigma_\pm(\varepsilon) = \sum_\alpha \left| \langle 0 | \boldsymbol{\mu}^{\text{tot}} | e_\pm^{(\alpha)} \rangle \right|^2 L_H\left(\varepsilon - \varepsilon_\pm^{(\alpha)}\right)$, $\boldsymbol{\mu}^{\text{tot}} = \mu_{eg}\left(\hat{\boldsymbol{d}}_A + \hat{\boldsymbol{d}}_B\right)$ is the collective EDTM and $L_H(\varepsilon) = \frac{1}{2}\Gamma_H / \left[\varepsilon^2 + \left(\frac{1}{2}\Gamma_H\right)^2\right]$ is a Lorentzian homogeneous lineshape with eigen-energy, $\varepsilon_\pm^{(\alpha)}$, and FWHM line width, $\Gamma_H = 186$ cm$^{-1}$. Similarly, the dimer homogeneous CD spectrum is the sum of symmetric and anti-symmetric rotational strengths [47]

$$CD_H^D(\varepsilon) = \sum_\alpha \left[ RS_+^{(\alpha)} L_H\left(\varepsilon - \varepsilon_+^{(\alpha)}\right) + RS_-^{(\alpha)} L_H\left(\varepsilon - \varepsilon_-^{(\alpha)}\right) \right] \quad (6)$$

where $RS_\pm^{(\alpha)} = \frac{\varepsilon_{eg}}{4\hbar c} \left\langle 0 \middle| \hat{d}_A \middle| e_\pm^{(\alpha)} \right\rangle \times \left\langle e_\pm^{(\alpha)} \middle| \hat{d}_B \middle| 0 \right\rangle \cdot \boldsymbol{R}_{AB}$. In the above expressions, the ground electronic-vibrational state of the $AB$ dimer is $|0\rangle = |n_A = 0, n_B = 0\rangle |gg\rangle$.

In solution, the (iCy3)$_2$ dimer conformation may vary from molecule to molecule due to local (and position-specific) DNA 'breathing' fluctuations. To account for spectral inhomogeneity, the homogeneous dimer absorption and CD line shapes are convolved with an inhomogeneous distribution function, $G\left(\varepsilon_\pm^{(\alpha)}\right) = (2\pi\sigma_I^2)^{-\frac{1}{2}} \exp\left[-\left(\varepsilon_\pm^{(\alpha)} - \bar{\varepsilon}_\pm^{(\alpha)}\right)^2 / 2\sigma_I^2\right]$, which is centered at the average transition energy, $\bar{\varepsilon}_\pm^{(\alpha)}$, and has standard deviation, $\sigma_I$. The final expressions for the absorption and CD spectra are given by the convolution integrals:

$$\sigma_I^D(\varepsilon) = \int_{-\infty}^{\infty} \sigma_H^D(\varepsilon - \varepsilon') G(\varepsilon') d\varepsilon' \quad (7)$$

$$CD_I^D(\varepsilon) = \int_{-\infty}^{\infty} CD_H^D(\varepsilon - \varepsilon') G(\varepsilon') d\varepsilon' \quad (8)$$

According to the H-F model, the simulated absorption and CD spectra of the (iCy3)$_2$ dimer-labeled ss-dsDNA constructs depend on, in addition to the monomer Hamiltonian parameters (*i*) – (*iv*) discussed above, (*v*) the electrostatic coupling, $J$, and (*vi*) the spectral inhomogeneity parameter, $\sigma_I$. The value of $J$ depends on structural parameters that specify the conformation of the (iCy3)$_2$ dimer while the value of $\sigma_I$ depends on the conformational disorder, which can be determined experimentally using two-dimensional fluorescence spectroscopy (2DFS), as discussed in Sect. III.B below.

The electrostatic coupling, $J$, can be calculated using quantum chemical models of the Coulomb interaction between the iCy3 monomer electronic transition density [25, 42, 43].

$$J = \frac{1}{4\pi\epsilon} \int_{-\infty}^{\infty} d\boldsymbol{r}_A \int_{-\infty}^{\infty} d\boldsymbol{r}_B \frac{\rho_A^{ge}(\boldsymbol{r}_A)\rho_B^{eg}(\boldsymbol{r}_B)}{|\boldsymbol{R}_{AB}|} \qquad (9)$$

where $\boldsymbol{R}_{AB} = \boldsymbol{r}_B - \boldsymbol{r}_A$ is the inter-chromophore center-of-mass separation vector, $\rho_M^{ge}(\boldsymbol{r}_M) = \langle g|_M \hat{\rho}(\boldsymbol{r}_M)|e\rangle_M$ is the monomer transition density matrix element, and $\epsilon$ is the electric permittivity. The ability of these calculations to determine conformational parameters of (iCy3)$_2$ dimer-labeled ss-dsDNA constructs by analyzing experimental optical spectra has been assessed by comparing transition density models of varying levels of approximation [25]. In the point-dipole (PD) model, the electrostatic coupling is given by

$$J^{PD} = \frac{|\mu_{eg}|^2}{4\pi\epsilon|R_{AB}|^3} [\hat{d}_A \cdot \hat{d}_B - 3(\hat{d}_A \cdot \hat{R}_{AB})(\hat{R}_{AB} \cdot \hat{d}_B)] \qquad (10)$$

The PD model assumes that the transition density can be approximated as a point-dipole located at the molecular center-of-mass, which is an unrealistic approximation for intermolecular separations smaller than the iCy3 monomer dimension. The finite dimension of the molecule can be accounted for by using the 'extended dipole' (ED) model, which represents the transition density as a line segment that is oriented parallel to the monomer EDTM with equal and opposite charges on its endpoints [48, 49].

$$J^{ED} = \frac{|\mu_{eg}|^2}{4\pi\varepsilon l^2} \left[ \frac{1}{R_{AB}^{++}} - \frac{1}{R_{AB}^{-+}} - \frac{1}{R_{AB}^{+-}} + \frac{1}{R_{AB}^{--}} \right] \qquad (11)$$

Equation (11) approximates the transition density for each monomer as two separate point charges of equal magnitude ($q$) and opposite sign, which lie separated by the distance $l$ such that $ql = \mu_{eg}$. The distances between point charges are given by $R_{AB}^{\pm\pm} = [R_{AB} \pm l(\hat{d}_A - \hat{d}_B)/2]$ and $R_{AB}^{\mp\pm} = [R_{AB} \mp l(\hat{d}_A + \hat{d}_B)/2]$. The three-dimensional atomic structure of the iCy3 chromophore can be accounted for by applying an atomistically-detailed electrostatic coupling model that is based on 'transition charges' (TQs) derived from *ab initio* quantum chemical calculations. In the TQ model, the electrostatic interaction is approximated as a discrete sum over individual transition electrostatic potential charges ($q_i$), which are assigned to the centers of each atomic coordinate of the monomer [42].

$$J^{TQ} = \frac{1}{4\pi\epsilon}\sum_a^{N_A}\sum_b^{N_B}\frac{q_a q_b}{|r_a^A - r_b^B|} \quad (12)$$

In Eq. (12), $\boldsymbol{r}_i^M$ is the position of the $i^{\text{th}}$ atom of chromophore $M$ ($\in \{A, B\}$) and $N_M$ is the number of atoms in chromophore $M$.

A minimal set of structural parameters are used to characterize the conformations of (iCy3)$_2$ dimer-labeled ss-dsDNA constructs. For the PD and ED models, the conformation is fully specified by the 'tilt,' $\theta_{AB}$, 'twist,' $\phi_{AB}$, and 'inter-chromophore separation,' $R_{AB}$ (Figs. 5B and 5C). However, for TQ model calculations, which include higher dimensionality, additional structural parameters are needed to distinguish between relative orientations and translational displacements of the iCy3 monomers. These are the 'roll' angle, $\eta_{AB}$, to specify the degree of chromophore stacking, and the vertical 'shear,' $\delta_{AB}$, and the horizontal 'shift,' $\xi_{AB}$, displacements [25]. For a specific set of conformational parameters, and depending on the choice of coupling model, the value of $J$ and simulated optical spectra are calculated and compared to experimental data. 'Optimized values' for the conformational parameters are thus obtained by applying the above procedure iteratively while rejecting sterically overlapping dimer conformations using a multi-parameter optimization procedure.

The above procedure provides nearly identical results for the three different coupling models in terms of the twist angle, $\phi_{AB}$, tilt, $\theta_{AB}$, roll, $\eta_{AB}$ and inter-chromophore separation $R_{AB}$. In principle, information about the roll angle, $\eta_{AB}$, is only available using the atomistically-detailed TQ model [25]. However, by systematically rejecting sterically overlapping dimer conformations (approximating the van der Waals radii of the major component atoms of the iCy3 monomers as $r_w \sim 1.5$Å), corroborating information between the three models for all structural parameters are obtained. Thus, the reliability of the above approach for determining structural information from optical spectra is established.

In Fig. 6 are shown comparisons between experimental and simulated room temperature optical spectra of (iCy3)$_2$ dimer-labeled ss-dsDNA fork constructs labeled at the +2, +1, -1 and -2 positions. The experimental CD and absorption spectra (green curves) are shown overlaid with simulated spectra resulting from the H-F model analysis. Optimized values for the corresponding structural parameters are presented in the insets. The results indicate that the electrostatic coupling, $J$, undergoes a sign inversion as the (iCy3)$_2$ dimer probe position is changed from the +1 to -1 position across the ss-dsDNA junction. This is accompanied by non-continuous changes of the local conformational coordinates: the interchromophore separation $R_{AB}$ [6.2 Å (+2), 7.3 Å (+1), 7.0 Å (-1) and 7.5 Å (-2)], the twist angle $\phi_{AB}$ [85º (+2), 85º (+1), 96º (-1) and 98º (-2)], and the tilt angle $\theta_{AB}$ [1º (+2), 30º (+1), 25º (-1) and 15 (-2)]. The sign inversion of the electrostatic coupling between the +1 and -1 positions indicates that the conformation of the (iCy3)$_2$ dimer probe, and presumably that of the sugar-phosphate backbones labeled at these sites, changes from right-handed to left-handed. This change of handedness is correlated to a change of the twist angle $\phi_{AB}$ from values that are less than 90º (right-handed) to values that are greater than 90º (left-handed). In addition, the tilt angle $\theta_{AB}$ undergoes an abrupt increase from 1º to 30º between the +2 and +1 positions, followed by a decrease to ~25º at the -1 position, and again to ~15º at the -2 position. This indicates that the local conformation of the sugar-phosphate backbones within the ss-dsDNA fork constructs undergo an abrupt loss of cylindrical symmetry at the interface between the +2 and +1 positions. However, some of the cylindrical symmetry is recovered at the -1 and -2 positions.

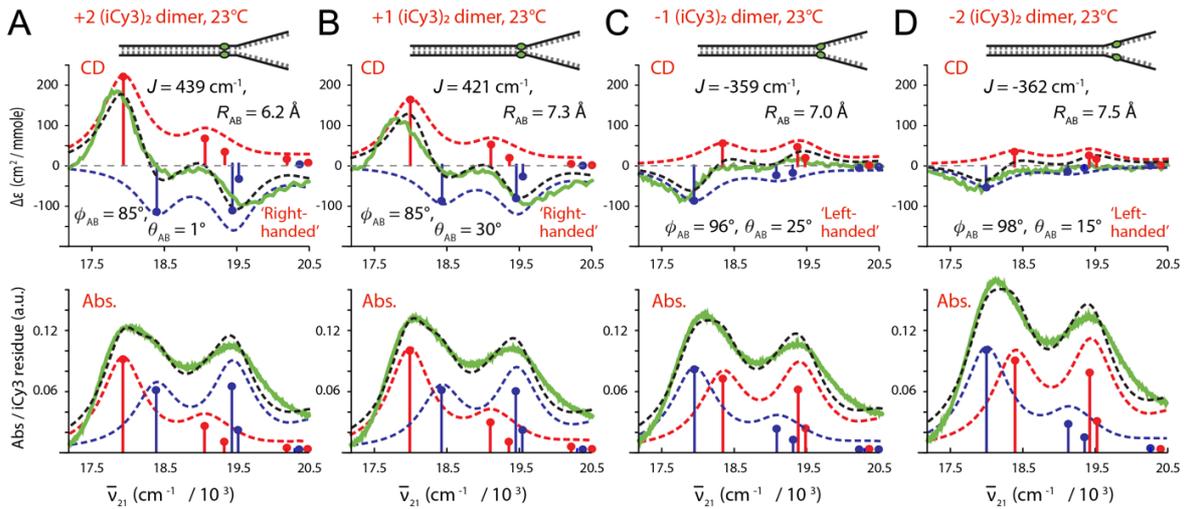

**Figure 6.** Experimental and simulated CD and absorption measurements performed at room temperature (25°C) for (iCy3)$_2$ dimer ss-dsDNA constructs as a function of probe labeling position. (*A*) +2; (*B*) +1; (*C*) -1; and (*D*) -2. Samples were prepared using 100 mM NaCl and 6 mM MgCl$_2$, and 10 mM Tris at pH 8.0. The experimental CD (top row) and absorption spectra (bottom row) are shown (in green) overlaid with vibronic spectral features (black dashed curves) obtained from optimized fits to the H-F model. The symmetric (+) and anti-symmetric (–) exciton state manifold contributions to the spectra are shown separately in blue and red, respectively. Values of the optimized parameters are shown in the insets of the corresponding panels. The mean local conformation changes abruptly from a right-handed to a left-handed twist symmetry as the label position is changed from the +1 to the -1 position. Figure adapted from Refs. [24, 25].

**III.B Adiabatic Franck-Condon models of iCy3 monomer and (iCy3)$_2$ dimers**

While 'effective mode' models for simulating the optical spectra of iCy3 monomer and (iCy3)$_2$ dimer-labeled DNA constructs are useful for their relative simplicity of implementation, more accurate computational approaches are also available to test the validity of such utilitarian models. Of particular interest is the Franck-Condon (FC) approach, which considers all the vibrational degrees of freedom of the iCy3 molecule rather than a single effective mode. The FC approach is particularly valuable for simulating the spectra of molecules with strong electron-vibrational coupling. Unlike the minimal state 'effective mode' models, the adiabatic FC approach is a full *ab initio* method that accounts for excited state processes and all normal modes on an equal footing [50, 51].

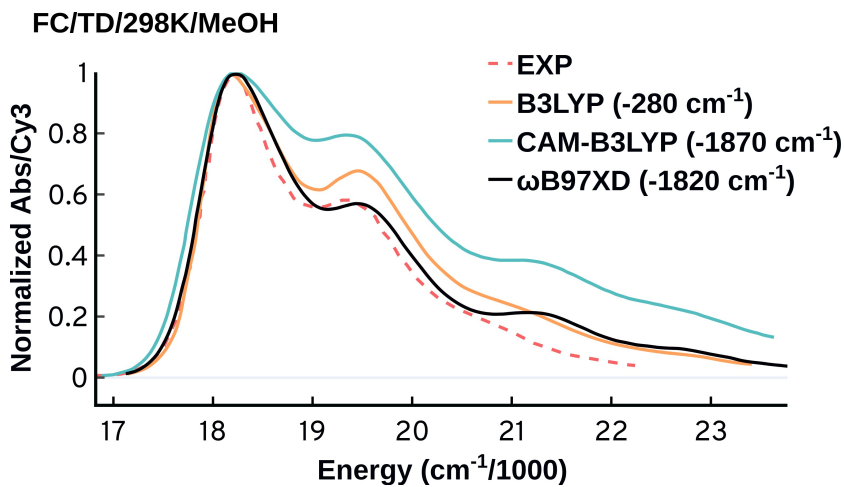

**Figure 7.** Time-dependent Franck-Condon (FC) density functional theory (DFT) calculations of the Cy3 monomer. Comparisons are shown between the B3LYP, CAM-B3LYP, and ωB97XD functionals at 298K. The calculations were performed using implicit methanol (MeOH) solvent via a polarizable continuum model. The calculated spectra were shifted to match the experimental absorption maximum with the shift value shown in parentheses. From Ref. [37].

Within the FC approach a harmonic potential is assumed for both the ground and excited states from which the normal modes are obtained. The probability of electronic transitions is calculated using the Franck-Condon factors, which describe the overlap between the vibrational wave functions of the ground and excited (final) electronic states at their respective equilibrium positions, as described by [50, 51]

$$\langle \Psi' | \mu | \Psi'' \rangle = \langle \chi_v' | \chi_v'' \rangle \langle \psi_e' | \mu | \psi_e'' \rangle \tag{13}$$

where $\Psi$ and $\psi_e$ are the total and electronic wavefunctions, respectively, of the initial ($''$) and final ($'$) states; $\chi_v$ is the vibrational wavefunction, and $\mu$ is the electronic transition dipole operator.[2] The FC approach has been used to simulate the absorption spectra of several members of the cyanine family with reasonable accuracy [37, 52]. The FC approach is highly sensitive to the choice of quantum mechanical method (*i.e.*, the choice of functional) used for the structural optimization and calculation of the spectrum. In Fig. 7 are shown comparisons between the experimental spectrum of the Cy3 chromophore in methanol and FC calculations using the B3LYP, CAM-B3LYP and the ωB97XD functionals, with the latter providing the best agreement with experiment.

The FC approach calculates the overlap terms between vibrational wavefunctions of all the normal modes, and thus provides detailed information about the vibrational excitations contributing to the different bands of the spectrum. In Fig. 8 are shown the vibronic excitations contributing to the Cy3 monomer spectrum obtained from the FC approach. The first (lowest energy) spectral band is dominated by low-frequency modes, *e.g.* mode 3 (with frequency = 35 cm$^{-1}$) and combined vibronic excitations from the same and other normal modes. The labels in Fig. 8 indicate the mode numbers of specific vibronic transitions with significant contributions to the spectrum. For example, 0-0 indicates the vertical FC transition without vibrational excitation, 0-3$^n$ indicates the n*th* FC transition (in order of increasing energy) that is coupled to mode 3, and 0-30$^1$3$^n$ indicates the FC combined transition, which is coupled to mode 30 through its anharmonic coupling to mode 3$^n$. The higher energy bands are dominated by combined vibrational excitations between normal mode number 3 and higher energy modes with frequencies ranging from 1,100 cm$^{-1}$ to 1,550 cm$^{-1}$. Thus, the broadly defined shape of the Cy3 monomer absorption spectrum – with 'effective mode' vibrational spacing ~1,100 cm$^{-1}$ – can be understood in terms of the direct coupling of the optical transition to mode 3, which in turn couples indirectly (anharmonically) to higher frequency modes. These findings emphasize the importance of accounting for all normal modes to accurately model and interpret the Cy3 spectrum. Moreover, they provide an understanding that justifies the application of minimal state 'effective mode' models for simulations of iCy3 monomer and (iCy3)$_2$ dimer-labeled DNA spectra, as discussed in Sect. III.A.

Full quantum mechanical modeling of the optical spectra of (iCy3)$_2$ dimer-labeled ss-dsDNA constructs is less straightforward than for the iCy3 monomer due to the need to optimize the minima of the ground and excited electronic state surfaces of such multi-subunit flexible molecules. As discussed in Sect. III.A, approximation methods for calculating the electrostatic coupling, $J$, within an $(iCy3)_2$ dimer can be obtained using a 'transition charge' TQ model derived from quantum chemical calculations [42]. The TQ approach was applied successfully to model the optical spectra of $(iCy3)_2$ dimer-labeled DNA constructs to obtain structural information [25]. Furthermore, the FC approach was used to accurately model a well-defined $(Cy3)_2$ dimer in which the Cy3 monomers were directly attached to each other using covalent linkers [43]. Such FC calculations provide detailed information about the electronic-vibrational properties of the $(Cy3)_2$ dimer spectral features, which cannot be accurately determined using minimal state 'effective mode' models.

---

[2] Note that in the current section we adopt a slightly different nomenclature for vibrational and electronic wavefunctions and transition dipole operators than in previous sections to be consistent with the cited literature.

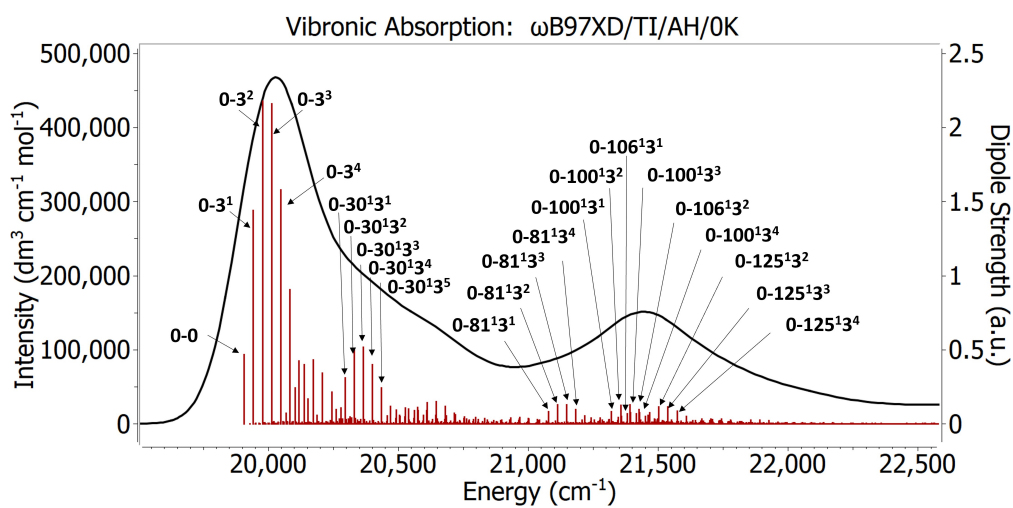

**Figure 8.** The vibronic spectrum of the Cy3 monomer obtained from structures optimized using the ωB97XD functional and the FC / Adiabatic Hessian (AH) approach at 0K. The red sticks correspond to the vibronic excitations. The labels on the sticks are selected transitions with significant contributions. TI: Time-independent. From Ref. [37].

**III.C Characterization of conformational disorder of exciton-coupled $(iCy3)_2$ dimer-labeled ss-dsDNA fork constructs by two-dimensional fluorescence spectroscopy (2DFS)**

While absorption and CD measurements can be used to determine the mean structural parameters of the $(iCy3)_2$ dimer probes within site-specifically labeled ss-dsDNA constructs, two-dimensional fluorescence spectroscopy (2DFS) provides additional information about the distributions of these parameters [24]. 2DFS is a fluorescence-detected Fourier transform (FT) optical method that is well suited to measure the electronic-vibrational structure of fluorescent molecular chromophores [53-57]. The underlying spectroscopic principles of 2DFS resemble those of 2D NMR [58, 59] and 2DIR [60, 61], with the latter providing structural and dynamic information about local vibrational modes in proteins [62, 63], nucleic acids [64, 65] and biomolecular hydration shells [66]. By virtue of its ability to sensitively monitor fluorescence against a dark background, 2DFS can detect relatively weak excited state population signals above the background of scattered excitation light. The signals detected by 2DFS on (iCy3)$_2$ dimer-labeled ss-dsDNA constructs directly monitor DNA backbone conformations and conformational disorder, which likely play a central role in protein recognition and binding events.

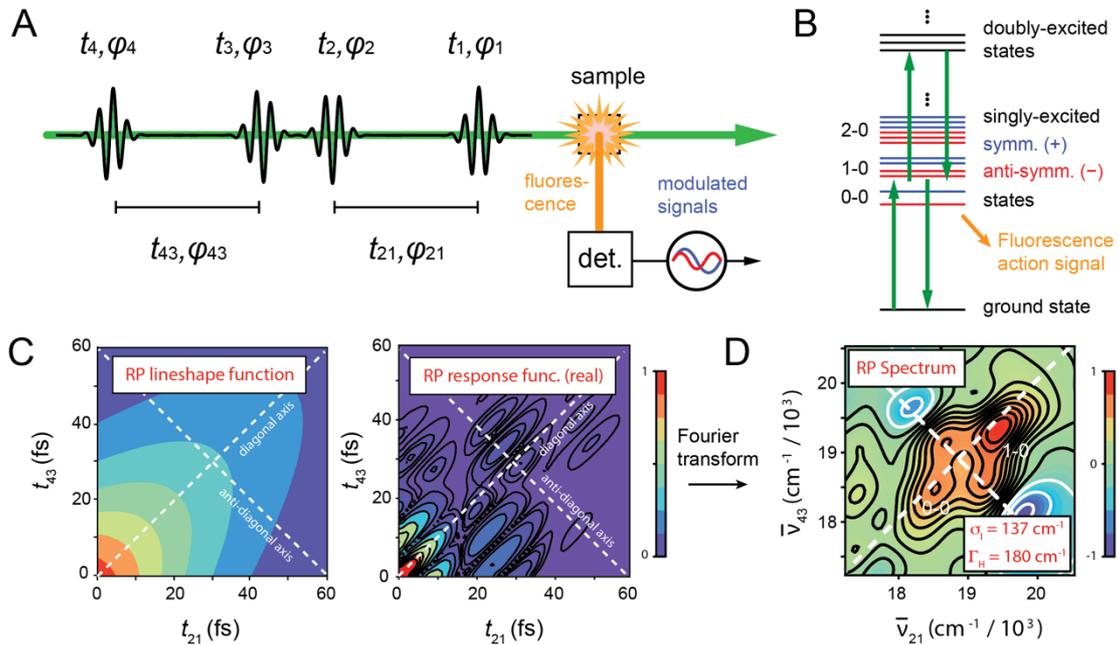

**Figure 9.** (*A*) The 2DFS instrument prepares a sequence of four collinear optical pulses with relative phases, $\varphi_{21}$ and $\varphi_{43}$, and time delays, $t_{21}$ and $t_{43}$. The ensuing fluorescence signals are detected as a function of the inter-pulse delays and phases. (*B*) Energy level diagram indicating the laser pulse energy and exciton-coupled (iCy3)$_2$ dimer state energies. Each laser pulse can induce a transition between ground, singly-excited and doubly-excited exciton states. Fluorescence is spectrally separated from scattered laser excitation light. (*C*) Example calculations of the real parts of the 2DFS rephasing (RP) line shape (left panel) and response (right panel) functions. These functions are displayed as two-dimensional contour plots with diagonal and anti-diagonal axes indicated as white dashed lines. The RP response function, $S_{RP}(t_{21}, t_{32} = 0, t_{43})$, shown in the right panel contains both the line shape function and the transition frequency phase factors. (*D*) The experimental RP 2D fluorescence spectrum, $\hat{S}_{RP}(\bar{\nu}_{21}, t_{32} = 0, \bar{\nu}_{43})$, of the +15 (iCy3)$_2$ dimer-labeled ss-dsDNA construct is calculated by performing a two-dimensional FT of the experimental response function with respect to the time variables $t_{21}$ and $t_{43}$. Comparisons between model calculations and experimental data determine the values of the homogeneous and inhomogeneous line shape parameters, $\Gamma_H$ and $\sigma_I$, respectively. Optimized values of $\Gamma_H$ and $\sigma_I$ are given in the inset. Figure adapted from Refs. [24, 41].

In 2DFS, a repeating sequence of four collinear optical pulses (see Fig. 9A) is generated using interferometric methods. The pulses are characterized by their inter-pulse delays, $t_{ij} = t_i - t_j$, and relative phases, $\varphi_{ij}$ [$i,j \in \{1,2,3,4\}$]. For a given measurement, the time delays, $t_{21}$, $t_{32}$ and $t_{43}$, are held fixed while the relative phases, $\varphi_{21}$, $\varphi_{32}$ and $\varphi_{43}$, are swept continuously at tens-of-kilohertz frequencies [53]. The laser is optically resonant with the symmetric and anti-symmetric excitons of the (iCy3)$_2$ dimer probes, such that each of the four pulses can excite a one-photon dipole-allowed optical transition between ground, singly-excited and doubly-excited states (see Fig. 9B). The resulting phase-dependent fluorescence signals are proportional to the nonlinear populations of molecular excited states, which can be detected separately using phase sensitive methods [67]. 'Rephasing' (RP) and 'non-rephasing' (NRP) signals are of specific interest, which have 'phase signatures' $\varphi_{43} + \varphi_{21}$ and $\varphi_{43} - \varphi_{21}$, respectively. The recorded signals are complex-valued response functions of the inter-pulse delays, $S_{RP}(t_{21}, t_{32} = 0, t_{43})$ and $S_{NRP}(t_{21}, t_{32} = 0, t_{43})$, which oscillate in time at the transition frequencies of the (iCy3)$_2$ dimer probe (see Fig. 9C) [24]. The RP response function contains the line shape factor $\exp[-\Gamma_H(t_{21} + t_{43})]\exp[-\sigma_I^2(t_{21} - t_{43})^2/2]$, which decays exponentially at the homogeneous dephasing rate along the diagonal axis ($t_{21} + t_{43}$) and as a Gaussian envelope with inhomogeneous decay constant $\sigma_I^2/2$ along the anti-diagonal axis ($t_{21} - t_{43}$). In contrast, the NRP response function contains the line shape factor $\exp[-\Gamma_H(t_{21} + t_{43})]\exp[-\sigma_I^2(t_{21} + t_{43})^2/2]$, which decays along the diagonal axis with rate constants that depend on both the homogeneous and inhomogeneous parameters [24]. Simultaneous comparisons of both RP and NRP simulated and experimental 2D spectra permits accurate determination of the homogeneous and inhomogeneous line shape parameters, $\Gamma_H$ and $\sigma_I$, respectively. FT of the response functions with respect to $t_{21}$ and $t_{43}$ yields the frequency-dependent RP and NRP spectra, $\hat{S}_{RP}(\bar{\nu}_{21}, t_{32} = 0, \bar{\nu}_{43})$ and $\hat{S}_{NRP}(\bar{\nu}_{21}, t_{32} = 0, \bar{\nu}_{43})$, respectively. An example experimental RP 2DFS spectrum of the +15 (iCy3)$_2$ dimer-labeled ss-dsDNA construct, and the resulting values for $\Gamma_H$ and $\sigma_I$, are shown in Fig. 9D.

In Fig. 10 are shown temperature-dependent measurements of the +15 (iCy3)$_2$ dimer-labeled ss-dsDNA construct, which includes the absorption, CD, and experimental and simulated RP 2DFS data [24]. At the lowest temperatures (5 – 23°C), the absorption and CD spectra exhibit, respectively, the intensity borrowing and bisignate lineshapes that are characteristic of exciton-coupled (iCy3)$_2$ dimer probes (Figs. 10A and 10B). The 2DFS data at these relatively low temperatures show well-separated peaks and cross-peaks, which indicate the presence of the spatially delocalized symmetric (+) and anti-symmetric (−) excitons (see discussion in Sect. III.A).

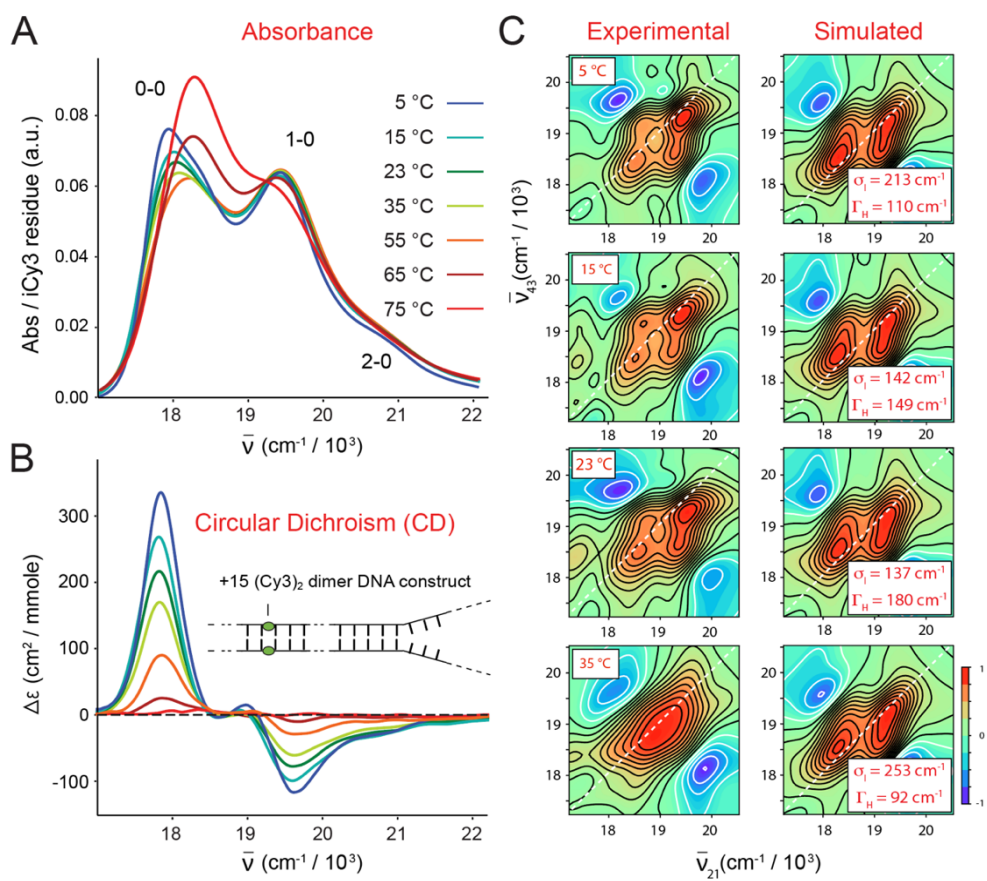

**Figure 10.** Melting behavior of the +15 (iCy3)$_2$ dimer-labeled ss-dsDNA construct. Temperature-dependent measurements of the (*A*) absorption, (*B*) CD, (*C*) experimental and simulated 2DFS data. Comparisons between model calculations and experimental data determine the values of the homogeneous and inhomogeneous line shape parameters, $\Gamma_H$ and $\sigma_I$, respectively. Optimized values of $\Gamma_H$ and $\sigma_I$ are given in the insets. Samples were prepared using 100 mM NaCl and 6 mM MgCl$_2$, and 10 mM Tris at pH 8.0. Figure adapted from Ref. [24].

As the temperature is elevated above room temperature towards the melting point ($T_m$ = ~65°C), the Watson-Crick (W-C) B-form structure of the DNA bases and sugar-phosphate backbones becomes locally unstable. The absorption and CD spectra change systematically with increasing temperature to reflect the reduced coupling strength of the (iCy3)$_2$ dimer probe due to the disruption of structural order. At the highest temperature for which the single strands of the DNA duplex have fully separated (75°C) the coupling strength is approximately zero such that the absorption and CD spectra resemble those of the iCy3 monomer substituted ss-dsDNA constructs. The 2DFS data reveal the sensitivity of this method to the presence of local conformational disorder that is 'sensed' by the (iCy3)$_2$ dimer probe deep within the duplex region (+15 position) of the ss-dsDNA construct. At low temperatures, the spectral features of the 2DFS data are well-defined. However, at elevated temperatures the spectral features become progressively less pronounced and the exciton-split diagonal features merge into a single diffuse feature (Fig. 10*C*).

Analyses of 2DFS line shape data provide temperature-dependent values for the homogeneous and inhomogeneous line shape parameters, $\Gamma_H$ and $\sigma_I$, respectively [24]. Such measurements show how conformational disorder of (iCy3)$_2$ dimer-labeled ss-dsDNA constructs – proportional to $\sigma_I$ – vary with temperature and probe-labeling position. In Fig. 11 are shown the results of such studies for the +15 'duplex' and the -1 'fork' (iCy3)$_2$ dimer-labeled ss-dsDNA construct. The values of $\Gamma_H$ and $\sigma_I$ are shown as blue- and teal-shaded regions, respectively. The optimized values are presented as points, and the shaded regions bounded by dashed lines indicate error bars.

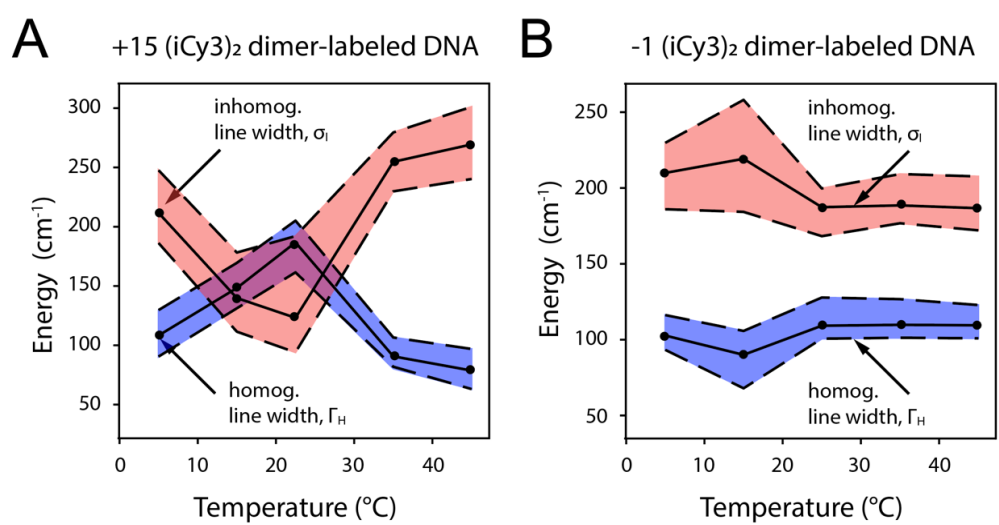

**Figure 11.** Optimized homogeneous ($\Gamma_H$, blue) and inhomogeneous ($\sigma_I$, teal) line width parameters as a function of temperature obtained from 2DFS line shape analyses. (*A*) +15 (iCy3)$_2$ dimer-labeled ss-dsDNA construct; (*B*) -1 (iCy3)$_2$ dimer-labeled ss-dsDNA construct. Shaded regions bounded by dashed lines indicate error bars determined from the analysis. The melting temperature, $T_m$, of the duplex regions of the DNA constructs is 65°C. Samples were prepared using 100 mM NaCl and 6 mM MgCl$_2$, and 10 mM Tris at pH 8.0. Figure adapted from Refs. [24, 41].

For the +15 'duplex' (iCy3)$_2$ dimer-labeled ss-dsDNA construct, the local conformation is minimally disordered at room temperature (see Fig. 11*A*). As the temperature is raised or lowered from room temperature (23°C), the local conformational disorder abruptly increases, while the interactions between the probes and the solvent environment (proportional to the homogeneous linewidth) decreases. This indicates that the W-C B-form conformation of the (iCy3)$_2$ dimer probes at local sites deep within the duplex region is a marginally stable structure at room temperature, which becomes unstable at temperatures just above or below room temperature. These findings are consistent with the notion that the room temperature stability of the local B-form conformation results from a nearly equal balance between opposing thermodynamic forces (entropy-enthalpy compensation), and that small departures from room temperature (in either the positive or negative directions) alter the free energy landscape and serve to populate non-B-form conformations [68-70]. Such a picture is sometimes invoked in 'DNA breathing and trapping models,' in which non-canonical local conformations of the DNA framework are transiently populated under 'physiological' conditions and function as activated states required for protein-DNA complex assembly [2, 5, 71-73].

For the -1 'fork' (iCy3)$_2$ dimer-labeled ss-dsDNA construct, in which the average local conformation of the sugar-phosphate backbones at room temperature is left-handed and relatively disordered (in comparison to the +15 position), increasing the temperature above 23°C does not further increase the local conformational disorder (see Fig. 11*B*). However, decreasing temperature below 23°C does lead to a significant increase of conformational disorder at the -1 position. These findings indicate that the onset of local conformational disorder at elevated and reduced temperatures observed for the +15 labeled construct is associated with 'breathing' fluctuations of the DNA duplex. For the -1 labeled construct, the (iCy3)$_2$ dimer probes 'sense' the local conformations of the bases and backbones at and near the ss-dsDNA junction, where the local W-C B-form conformations are less stable and more susceptible to 'breathing' fluctuations in comparison to positions deep within the duplex region. Thus, increasing temperature leads to dissociation of the DNA duplex near the ss-dsDNA junction, while decreasing temperature leads to trapping of disordered local conformations near the ss-dsDNA junction.

In Fig. 12 are shown the results of 2DFS studies of local conformational disorder in (iCy3)$_2$ dimer-labeled ss-dsDNA constructs at sites traversing the ss-dsDNA junction. Analyses of these data show that the local conformational disorder increases as the dimer probe position is changed from the +2 to the +1 position. However, the local conformational disorder slightly decreases as the position is changed from the +1 to -1 position, and again from the -1 to -2 position. Thus, the position-dependent behavior of the conformational disorder parameter, $\sigma_I$, reflect the same trends observed for the mean local conformational parameters, $\phi_{AB}$, $\theta_{AB}$ and $R_{AB}$, which were discussed in Sect. III.A and summarized in Fig. 6.

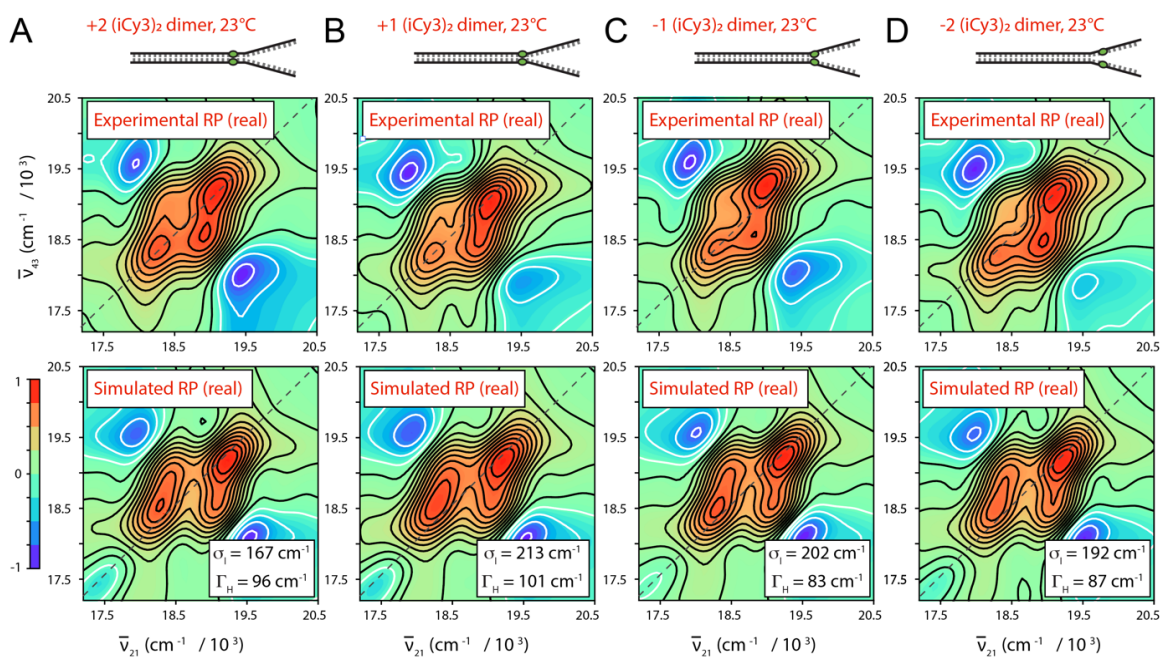

**Figure 12.** Experimental RP spectra (top row) are compared to the optimized simulated RP spectra (bottom row) at room temperature (23°C) for (iCy3)$_2$ dimer ss-dsDNA constructs as a function of probe labeling position. (*A*) +2; (*B*) +1; (*C*) -1; and (*D*) -2. Samples were prepared using 100 mM NaCl and 6 mM MgCl$_2$, and 10 mM Tris at pH 8.0. Simulated spectra are based on the structural parameters obtained from our optimization analyses of the CD and absorbance spectra. The resulting homogeneous and inhomogeneous line shape parameters ($\Gamma_H$ and $\sigma_I$, respectively) are given in the insets. The local conformational disorder increases as the probe labeling position is varied across the ss-dsDNA junction in the direction from the duplex to the single-stranded DNA region. Figure adapted from Refs. [24, 41].

In Fig. 13*A* is shown a summary of results of position-dependent studies of (iCy3)$_2$ dimer-labeled ss-dsDNA constructs [24]. These results suggest a detailed picture of the mean local conformations and conformational disorder of the bases and sugar-phosphate backbones at probe labeling sites at and near the ss-dsDNA junction. The average local conformation of the (iCy3)$_2$ dimer-probe changes systematically as the labeling site is varied from the +2 to the -2 positions. For positive integer positions the mean local conformation is right-handed (defined as exhibiting a mean twist angle, $\phi_{AB} < 90°$), and for negative integer positions the local conformation is left-handed ($\phi_{AB} > 90°$). Local conformations deep within the duplex region are cylindrically symmetric and minimally disordered. The disorder increases significantly for positive positions approaching the ss-dsDNA fork junction. At the +1 position, there is an abrupt loss of cylindrical symmetry, which coincides with an additional increase in conformational disorder. The left-handed conformations at the -1 and -2 positions exhibit somewhat smaller mean tilt angles and decreasing conformational disorder, suggesting that the peak perturbation to secondary structure within the ss-dsDNA fork junction occurs at the +1 position.

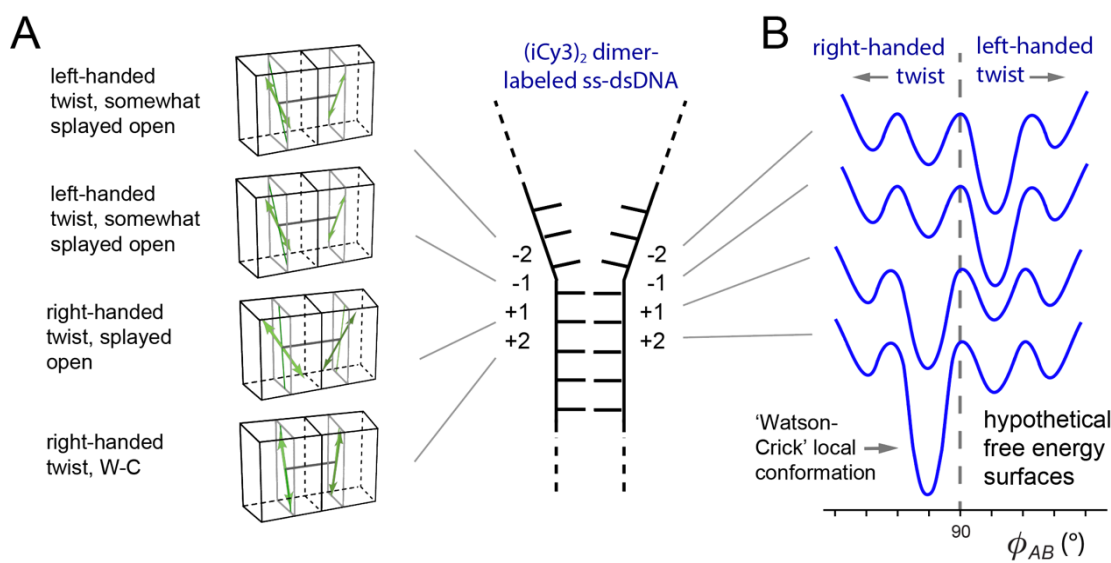

**Figure 13.** (*A*) Schematic of the position-dependent changes in mean local conformations of the (iCy3)$_2$ dimer-labeled ss-dsDNA constructs. (*B*) Hypothetical free energy landscape specific to the probe-labeling position. Individual surfaces are vertically offset for clarity. Figure adapted from Refs. [24, 41].

## III.D Studies of the equilibrium distributions of conformational macrostates and the kinetics of state-to-state interconversion by polarization-sweep single-molecule fluorescence (PS-SMF) microscopy

A significant finding of the studies of (iCy3)$_2$ dimer-labeled ss-dsDNA constructs described in Sect. III.C is that sites at and near the ss-dsDNA fork junction exhibit only moderate levels of structural disorder, suggesting that the distributions of conformational states at positions moving across the ss-dsDNA junction towards the single-stranded regions are relatively narrow. A possible interpretation of these results is that the (iCy3)$_2$ dimer probe, which is covalently linked to the sugar-phosphate backbones and bases immediately adjacent to the probe, is limited to only a small number of local right- or left-handed conformations with relative stabilities that depend on the probe labeling position relative to the ss-dsDNA junction. Figure 13*B* depicts hypothetical free energy surfaces to illustrate this point. Each surface contains four possible local conformations of the (iCy3)$_2$ dimer probe, which may exhibit either a right-handed or a left-handed twist. In this picture the free energy differences between the 'Watson-Crick' (B-form) ground state and the other (non-canonical) local conformations are sufficiently high to ensure that the Boltzmann-weighted distribution of available macrostates is dominated by the B-form structure. At the same time, the moderate free energies of activation allow for infrequent transitions between non-canonical local structures, which are present in trace amounts. As the probe-labeling position is varied across the ss-dsDNA junction towards the single-stranded region, one might expect the free energy surface to exhibit local minima with approximately the same coordinate values, but with relative stabilities and barrier heights shifted to reflect the presence of non-canonical structures observed in ensemble measurements [23-25].

To directly study the local conformational fluctuations – and the associated free energy surfaces – of (iCy3)$_2$ dimer-labeled ss-dsDNA fork constructs, a novel single-molecule optical method was developed. Polarization-sweep single-molecule fluorescence (PS-SMF) microscopy was designed to directly monitor the local conformational fluctuations of the (iCy3)$_2$ dimer probe as a function of its position relative to the ss-dsDNA fork junction [74, 75]. These measurements exploit the sensitivity of the conformation-dependent coherent coupling between the closely spaced monomers of the (iCy3)$_2$ dimer probe. While certain structural aspects of DNA 'breathing' in duplex DNA have previously been revealed using hydrogen exchange and chemical trapping studies [3, 4, 73, 76-78], PS-SMF experiments provide new and structure-specific information about the relative stabilities and transition barriers that mediate conformation fluctuations near the ss-dsDNA fork junctions characteristic of DNA replication complexes.

A schematic of the PS-SMF instrument is shown in Fig. 14. An (iCy3)$_2$ dimer-labeled ss-dsDNA fork construct is attached to the inner surface of a quartz sample chamber using a biotin/neutravidin linkage [79]. The sample is placed on the stage of an inverted TIRF (total internal reflection fluorescence) microscope. The optical source – a continuous wave (cw) laser with wavelength ~532 nm – is used to create an evanescent field, which propagates parallel to the slide surface with its electric field vector contained within a plane normal to the surface. The plane polarization state of the laser is prepared using a polarized Mach-Zehnder interferometer and quarter-wave plate, and the electric field vector is rotated continuously at the frequency 1 MHz by sweeping the phase of the interferometer using acousto-optic Bragg cells [74]. The symmetric (+) and anti-symmetric (−) excitons of the (iCy3)$_2$ dimer probe, which are oriented perpendicular to one another, are alternately excited as the laser polarization direction is rotated across the corresponding EDTMs. The modulated fluorescence from a single (iCy3)$_2$ dimer-labeled DNA construct is isolated through a pinhole and detected using an avalanche photodiode (APD). The signal is processed using the phase-tagged photon counting (PTPC) method [67].

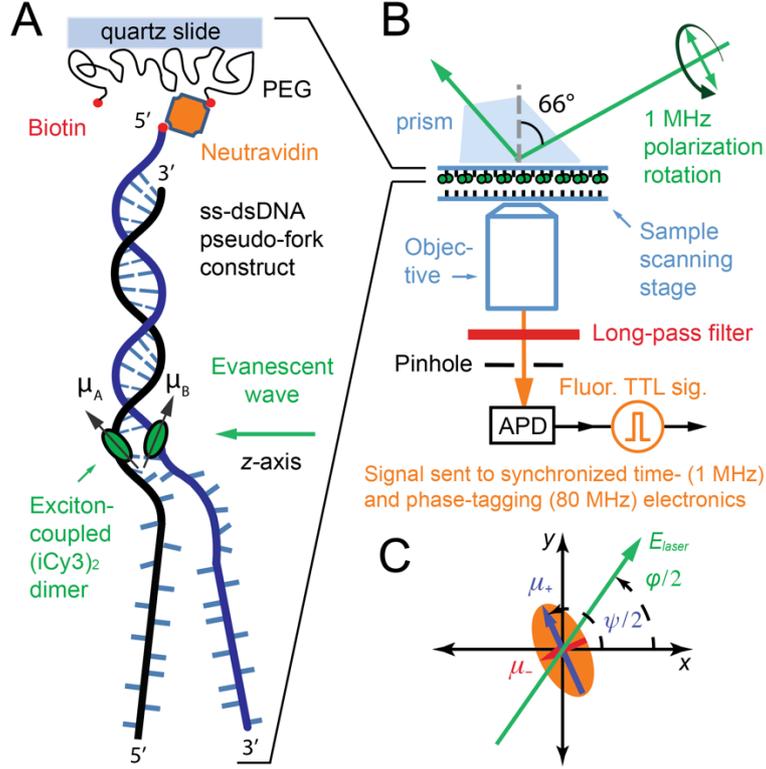

**Figure 14.** Schematic of PS-SMF microscopy setup. (*A*) (iCy3)$_2$ dimer-labeled ss-dsDNA constructs are attached to a microscope slide using biotin/neutravidin linkers. (*B*) A total internal reflection fluorescence (TIRF) illumination geometry is used to preferentially excite molecules attached near the liquid-solid interface. The cw 532 nm laser generates an evanescent wave that propagates parallel to the sample surface with electric field vector perpendicular to the surface. The plane-polarized electric field vector is rotated continuously at a frequency of 1 MHz by sweeping the phase of an interferometer. (*C*) The symmetric and anti-symmetric EDTMs of the (iCy3)$_2$ dimer probe (labeled $\boldsymbol{\mu}_\pm$, and shown as blue and red vectors, respectively) define major and minor axes of a 'polarization ellipse' in the transverse cross-sectional area of the incident laser beam. The angles $\psi/2$ and $\varphi/2$ define the orientations of the (iCy3)$_2$ dimer probe and the laser polarization vector, respectively. Reproduced from Ref. [74].

The PS-SMF signal has modulation amplitude that depends on the phase of the polarization rotation cycle. The signal modulation amplitude is termed the *visibility*, $v$, which is defined [74]:

$$v = [|\mu_+|^2 - |\mu_-|^2]/[|\mu_+|^2 + |\mu_-|^2] \tag{14}$$

In Eq. (14), $|\mu_\pm|^2 = |\boldsymbol{E}\cdot\boldsymbol{\mu}_\pm|^2 = |A(\omega)|^2 \sum_\alpha \int d\omega \left|\mu_\pm^{(\alpha)}(\omega)\right|^2$, are the intensities associated with the symmetric (+) and anti-symmetric (−) excitons, $\boldsymbol{E}$ is the plane-polarized electric field vector of the laser, $A(\omega)$ is the (narrow) spectral envelope, $\boldsymbol{\mu}_\pm^{(\alpha)} = \left\langle 0 \middle| \boldsymbol{\mu}^{\text{tot}} \middle| e_\pm^{(\alpha)} \right\rangle$ are the

symmetric and anti-symmetric EDTMs between electronic ground and excited states of the (iCy3)$_2$ dimer, and $\boldsymbol{\mu}^{\text{tot}} = \boldsymbol{\mu}_+ + \boldsymbol{\mu}_-$, as described in Sect. III.A.

As illustrated in Figs. 5*B* and 5*C*, the magnitudes of the $\boldsymbol{\mu}_\pm$ EDTMs depend sensitively on the local conformation of the (iCy3)$_2$ dimer. The orthogonal EDTMs define a 'polarization ellipse' in the cross-sectional area in which the laser beam projects onto the molecular frame, as shown in Fig. 14*C*. Thus, the visibility, $v$, is directly related to the instantaneous local conformation of the (iCy3)$_2$ dimer probe within the ss-dsDNA construct, which can be monitored on microsecond and longer time scales.

Studies of the local conformational fluctuations of (iCy3)$_2$ dimer-labeled ss-dsDNA fork constructs were recently carried out by monitoring time-dependent trajectories of the PS-SMF signal visibility, $v(t)$ [74]. The data were analyzed by constructing from a time series of PS-SMF measurements ensemble-averaged probability distribution functions (PDFs), $P(v)$, and the two-point and three-point time-correlation functions (TCFs), $\bar{C}^{(2)}(\tau)$ and $\bar{C}^{(3)}(\tau_1, \tau_2)$, respectively [21, 74, 80, 81]. The two-point TCF is the average product of two successive measurements separated by the interval $\tau$:

$$\bar{C}^{(2)}(\tau) = \langle \delta v(\tau) \delta v(0) \rangle \qquad (15)$$

The angle brackets in Eq. (14) indicate a running average over all possible initial measurement times, and $\delta v(t) = v(t) - \langle v \rangle$ is the signal fluctuation about the mean value, $\langle v \rangle$. The two-point TCF, $\bar{C}^{(2)}$, contains information about the number of quasi-stable macrostates of the system, the mean visibility of each macrostate, and the characteristic time scales of interconversion between macrostates [80]. The three-point TCF is the time-averaged product of three consecutive measurements separated by the intervals $\tau_1$ and $\tau_2$.

$$\bar{C}^{(3)}(\tau_1, \tau_2) = \langle \delta v(\tau_1 + \tau_2) \delta v(\tau_1) \delta v(0) \rangle \qquad (16)$$

The function $\bar{C}^{(3)}$ contains detailed kinetic information about the transition pathways that connect the macrostates of the system. This function is sensitive to the roles of intermediates, whose presence can facilitate or hinder successive transitions between macrostates [80]. Unlike even-moment TCFs, the three-point TCF, $\bar{C}^{(3)}$, is devoid of noise-related background contributions [82].

The above functions can be simulated using a kinetic network model, which assumes that the system interconverts between a finite number of Boltzmann-weighted macrostates, and that the rates of interconversion between macrostates are related to Arrhenius activation barriers. The minimum number of macrostates needed to fully describe PS-SMF data is equal to the number of decay components exhibited by the two-point TCF plus one. By comparing experimental results to the predictions of the kinetic network model, it is possible to extract the relative stabilities and activation barriers between distinct macrostates, which may be used to parameterize the free energy surfaces associated with the local environment of DNA bases and sugar-phosphate backbones 'sensed' by the (iCy3)$_2$ dimer probes.

In Fig. 15 are shown examples of PS-SMF data for the +1 (iCy3)$_2$ dimer-labeled ss-dsDNA fork construct. The mean signal intensity for these measurements was ~7,500 s$^{-1}$. From the PTPC data stream the signal visibility, $v(t)$, was determined using an integration period of ~100 ms (see Fig. 15$A$). The visibility trajectory undergoes discontinuous transitions between a small number of discrete values within the range $0 < v < 0.4$ (dashed lines are guides to the eye). The visibility is a direct measure of the conformational changes of the (iCy3)$_2$ dimer probe. Control measurements of iCy3 monomer-labeled DNA constructs do not exhibit discontinuous transitions, in contrast to (iCy3)$_2$ dimer-labeled ss-dsDNA constructs [74]. From ~40 individual PS-SMF measurements, each of ~30 s in duration, the ensemble-averaged functions were constructed. In Fig. 15$B$ is shown the probability distribution function (PDF), which was constructed using an integration period ~100 ms. The PDF exhibits multiple 'low visibility' and 'high visibility' features. There is a prominent 'low visibility' feature in the range $0 \lesssim v \lesssim 0.1$ in addition to sparsely populated 'high visibility' features in the range $0.1 \lesssim v \lesssim 0.4$. This broad distribution of visibility features indicates the presence of both stable and thermally activated local conformational macrostates of the (iCy3)$_2$ dimer probes.

The dynamics of the system are reflected by the behaviors of the two-point and three-point TCFs of the signal visibility. In Fig. 15$C$ is shown the two-point TCF, which was constructed using an integration period of ~250 $\mu$s. The two-point TCF is plotted alongside a model fit to a tri-exponential function, with time constants indicated in the figure. There are three decay components at $t_1 = 0.3$ ms, $t_2 = 3.7$ ms and $t_3 = 173$ ms. The presence of three well-separated decay components indicates that the underlying dynamics of the system must involve at least four quasi-stable macrostates in dynamic equilibrium. Figure 15$D$ shows the corresponding three-point TCF, which was constructed using an integration period of ~10 ms. The three-point TCF contains detailed kinetic information about the transition pathways that connect the macrostates of the system. The three-point TCF is sensitive to the roles of pathway intermediates, whose presence can facilitate or hinder successive transitions between macrostates.

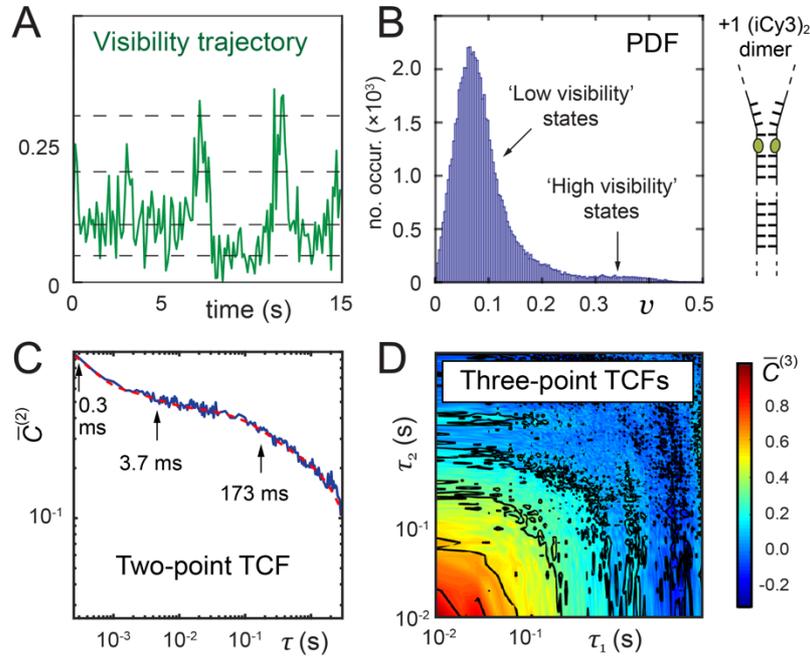

**Figure 15.** Sample experimental data from PS-SMF measurements of +1 (iCy3)$_2$ dimer-labeled ss-dsDNA construct in 100 mM NaCl, 6 mM MgCl$_2$, and 10 mM Tris at pH 8.0. (*A*) The first 15 seconds of a PS-SMF visibility trajectory, $v(t)$, is shown using a bin period of 100 ms. The mean signal intensity was ~7,500 s$^{-1}$. Horizontal dashed lines are guides to the eye to indicate the presence of multiple discrete conformational states. (*B*) Normalized probability distribution function (PDF) of visibility values, $P(v)$, using a bin period of 100 ms. (*C*) Two-point time-correlation function (TCF) is well modeled as a weighted sum of three exponentially decaying terms (red curve) with decay constants that are well-separated in time. (*D*) Three-point TCF is shown as a two-dimensional contour plot. Adapted from Ref. [75].

### III.E Studies of local DNA 'breathing' of +1, -1 and -2 (iCy3)$_2$ dimer-labeled ss-dsDNA fork constructs

In Fig. 16 are shown PS-SMF results for the +1, -1 and -2 (iCy3)$_2$ dimer-labeled ss-dsDNA constructs, which were obtained at room temperature (23°C) and 'physiological' buffer salt conditions ([NaCl] = 100 mM, [MgCl$_2$] = 6 mM). The left column shows the visibility PDFs, $P(v)$, the middle column shows the two-point TCFs, $\bar{C}^{(2)}(\tau)$, which are overlaid with multi-exponential fits, and the right column shows contour plots of the three-point TCFs, $\bar{C}^{(3)}(\tau_1, \tau_2)$. Like the PDF of the +1 (iCy3)$_2$ dimer-labeled ss-dsDNA construct (discussed above and shown in Fig. 16*A*), the PDFs of the -1 and -2 dimer-labeled constructs exhibit a major 'low visibility' feature in the region $0 \lesssim v \lesssim 0.1$, and minor 'high visibility' features in the region $0.1 \lesssim v \lesssim 0.4$. The relative weights of the 'high visibility' features depend on (iCy3)$_2$ dimer probe labeling position relative to the ss-dsDNA fork junction, with 'high visibility' features notably less prominent for the -1 construct than for the +1 and -2 constructs. Furthermore, the PDF of the -2 construct is significantly different from that of the +1 construct, as expected given the differential stabilities of stacked bases within dsDNA versus ssDNA regions spanning the ss-dsDNA fork junction.

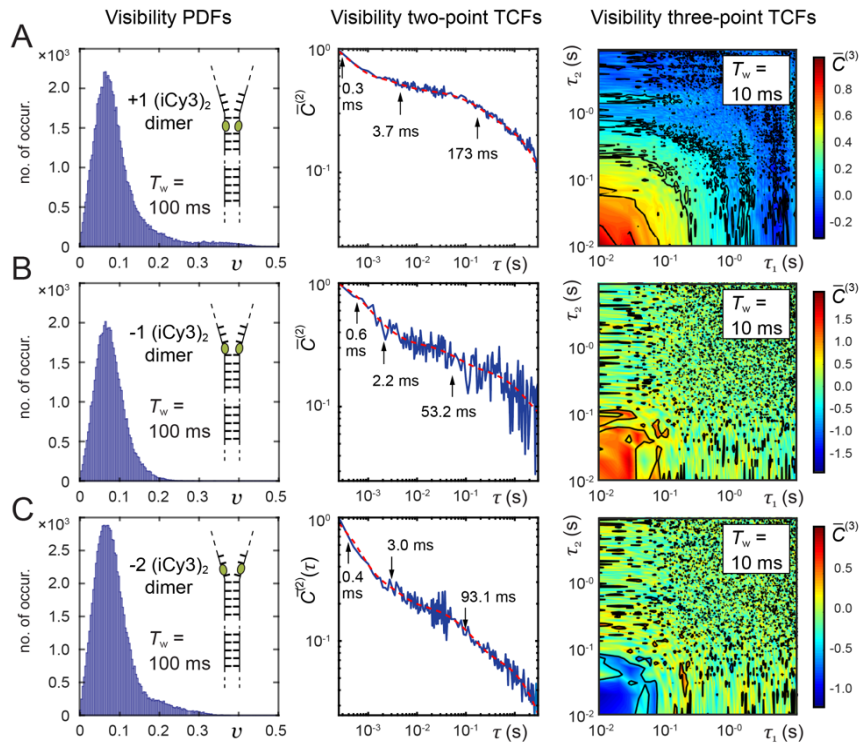

**Figure 16.** Probability distribution functions (PDFs, left column), two-point time-correlation functions (TCFs, middle column) and three-point TCFs (left column) of the PS-SMF visibility for the (**A**) +1, (**B**) -1 and (**C**) -2 (iCy3)$_2$ dimer-labeled ss-dsDNA fork constructs. The integration periods, $T_w$, used for the PDFs and three-point TCFs are indicated in the insets. The two-point TCFs were stitched together over the range 250 μs – 25 ms using $T_w$ = 250 μs, and from 25 ms – 2.5 s using $T_w$ = 10 ms. The two-point TCFs are shown overlaid with multi-exponential fits (dashed red curves). The three-point TCFs are plotted as two-dimensional contour diagrams. Experiments were performed at room temperature (23°C) in 100 mM NaCl, 6 mM MgCl$_2$, and 10 mM Tris at pH 8.0. Adapted from Ref. [74].

The two-point and three-point TCFs shown in Fig. 16 characterize the conformational dynamics of the (iCy3)$_2$ dimer-labeled ss-dsDNA constructs as a function of probe labeling position. The two-point TCFs are shown overlaid with multi-exponential fits (dashed red curves). In each case, the four decay components are well-separated in time: $t_1$ ~0.3 – 0.6 ms, $t_2$ ~2 – 4 ms, $t_3$ ~50 – 200 ms and $t_4$ ~1 – 2 s. The seconds-long decay, $t_4$, was determined to be due to mechanical room vibrations. Thus, there are three relevant, well-separated decay components present in the +1, -1 and -2 (iCy3)$_2$ dimer-labeled ss-dsDNA constructs under 'physiological' buffer salt conditions, suggesting the presence of four quasi-stable macrostates in thermal equilibrium.

The +1 (iCy3)$_2$ dimer-labeled ss-dsDNA construct exhibits the slowest relaxation dynamics of the three constructs examined in Fig. 16. As the position of the dimer probe is varied across the ss-dsDNA junction from +1 to -1, and again from -1 to -2, the relaxation dynamics become faster. These observations are consistent with the notion that the conformational fluctuations of the (iCy3)$_2$ dimer probe depend on the stabilities and dynamics of the DNA bases and sugar-phosphate backbones immediately adjacent to the probe. The W-C base pairs within the duplex side of the ss-dsDNA junction are largely stacked, while the stacking interactions between bases on the ssDNA side of the junction are much weaker due to the lack of complementary base pairing. A noteworthy feature of the three-point TCFs, $\bar{C}^{(3)}$, is that they exhibit positive amplitude for the +1 and -1 constructs, and negative amplitude for the -2 construct. These observations suggest that the prevailing kinetic pathways of state-to-state interconversion for the +1 and -1 constructs are significantly different from that of the -2 construct.

The results summarized in Fig. 16 indicate that the bases and sugar-phosphate backbones sensed by the (iCy3)$_2$ dimer probes can adopt four quasi-stable local conformations, whose relative stabilities and transition state barriers depend on position relative to the ss-dsDNA fork junction. Under 'physiological' buffer salt conditions, the +1 (iCy3)$_2$ ss-dsDNA construct exhibits a relatively broad distribution of local base and backbone conformations within the duplex region of the ss-dsDNA junction. The conformational macrostates of the +1 (iCy3)$_2$ ss-dsDNA construct undergo the slowest dynamics of the three positions investigated, indicating that the most thermodynamically stable states within the duplex side of the fork junction are also mechanically stable with relatively long population lifetimes. In comparison, the -2 (iCy3)$_2$ ss-dsDNA construct exhibits significantly faster dynamics and a distinctly different distribution of conformational macrostates, indicating that the thermodynamically favored conformations of bases and sugar-phosphate backbones within the ssDNA region of the fork junction are distinct from the duplex and mechanically unstable.

Additional insights about the energetics of the local conformational macrostates at these key positions across the ss-dsDNA fork junction were obtained from PS-SMF studies performed at varying salt concentrations [75]. For the +1 (iCy3)$_2$ ss-dsDNA construct, increasing or decreasing salt concentrations relative to 'physiological' concentrations shifted the equilibrium distribution to favor macrostates that are mechanically unstable with relatively low transition barriers. For the -2 (iCy3)$_2$ ss-dsDNA construct, varying salt concentration had the opposite effect, shifting the equilibrium distribution to favor macrostates that are mechanically stable with relatively high transition barriers. These results indicate that destabilization of the ss-dsDNA junction by varying salt concentrations leads to compensatory effects on the thermodynamic and mechanical stabilities of local conformational macrostates within the duplex and single-stranded DNA regions.

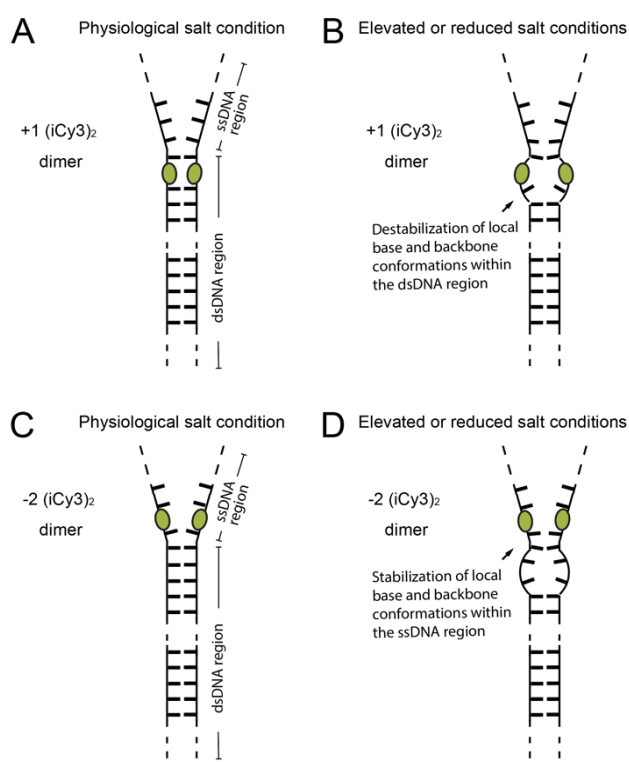

**Figure 17.** Schematic diagram illustrating the hypothesized mechanism of salt-induced instability of (iCy3)$_2$ dimer-labeled ss-dsDNA fork constructs labeled at (panel *A* and *B*) the +1 position and (panel *C* and *D*) the -2 position. Left column diagrams (panels *A* and *C*) represent the situation at 'physiological' salt conditions, and diagrams in the right column (panels *B* and *D*) represent the hypothesized mechanisms at elevated or reduced salt conditions. Adapted from Ref. [74].

In Fig. 17 is shown a hypothetical mechanism to account for the PS-SMF results discussed above. For the +1 (iCy3)$_2$ ss-dsDNA construct, the local base and backbone conformations adjacent to the dimer probes are within the DNA duplex region of the fork junction. At 'physiological' salt concentrations, the majority of conformational macrostates sensed by the probe are dominated by W-C base stacking, and these conformations are mechanically stable. Increasing or decreasing salt concentration leads to disruption of the local

W-C conformations within the duplex region and a reduction of mechanical stability. For the -2 (iCy3)$_2$ ss-dsDNA construct, the local base and backbone conformations adjacent to the dimer probes are within the ssDNA region. At 'physiological' salt concentrations, the majority of conformational macrostates sensed by the dimer in the ssDNA region are dominated by unstacked base conformations, which are distinct from the duplex region and mechanically unstable. Additional evidence for the presence of unstacked and dynamically labile base conformations within the ssDNA region of oligo(dT)$_{15}$ tails of ss-dsDNA fork constructs was determined by microsecond-resolved single-molecule FRET experiments [21] and corroborated by small-angle x-ray scattering experiments [83, 84]. At elevated or reduced salt concentrations, the distribution of conformational macrostates sensed by the dimer probes within the ssDNA region of the fork junction is shifted to mechanically stable base-stacked conformations. This hypothetical mechanism is consistent with the findings of previous ensemble studies of (iCy3)$_2$ dimer labeled ss-dsDNA fork constructs [24] and are further interpreted using kinetic network modeling approaches in a subsequent paper.

**III.F Inference of conformational macrostates of (iCy3)$_2$ dimer-labeled ss-dsDNA constructs**

The ensemble and single-molecule studies discussed in Sects. III.A – III.E indicate that the local conformations of DNA bases and backbones immediately adjacent to the (iCy3)$_2$ dimer probes fluctuate in thermal equilibrium between four quasi-stable macrostates. The results of these studies suggest that the FESs that govern the equilibrium distributions and interconversion dynamics are sensitive to position relative to the ss-dsDNA junction, as illustrated in Fig. 13. At positive integer positions within the duplex side of the ss-dsDNA junction the distribution of macrostates is dominated by a right-handed conformation, while at negative integer positions within the single-stranded side of the junction the distribution is dominated by a left-handed conformation. Based on the above observations and considering prior studies of DNA 'breathing' using hydrogen exchange methods [3, 4, 76, 77], it is possible to hypothesize structural assignments to the four conformational macrostates of the system.

From the work of Printz *et al*. and McConnell *et al*., it was concluded that a local 'cooperatively exchanging' region of duplex DNA (consisting of a ~5 – 10 base-pair segment) can undergo thermally activated transitions to populate unstable, 'open' conformations [2]. In Fig. 18 are illustrated four conformational macrostates (labeled S$_1$ – S$_4$) in order of increasing visibility, which are consistent with the experimental observations described in previous sections. The lowest visibility state, $S_1$, represents the so-called 'propped open' macrostate, in which a hydrogen bond within an affected base-pair is disrupted (*i.e.*, held open) by the gain or loss of an inter-base hydrogen atom, resulting in two H-bond donors or two H-bond acceptors facing one another within a W-C base pair. The propped open state is right-handed and has stacked flanking bases. Thus, in the $S_1$ macrostate, the complementary hydrogen bonds of opposing bases are separated by relatively large distances (indicated by horizontal arrows), such that the twist angle $\phi_{AB}$ approaches 90° and the visibility approaches zero. $S_2$ represents the canonical W-C conformation, which is right-handed and has stacked flanking bases. The $S_2$ macrostate has right-handed twist angle $\phi_{AB} \lesssim 90°$ [25]. $S_3$ represents a left-handed conformation in which flanking bases are unstacked by relatively large distances (indicated by vertical arrows). The $S_3$ macrostate has a left-handed twist angle $\phi_{AB} > 90°$. The $S_4$ macrostate, like $S_3$, has its flanking bases unstacked, but here with a right-handed twist angle $\phi_{AB} < 90°$.

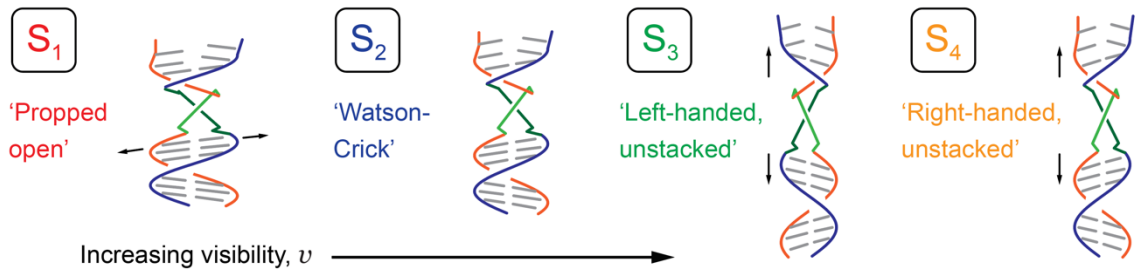

**Figure 18.** Conformational macrostates of (iCy3)$_2$ dimer-labeled ss-dsDNA constructs arranged in order of increasing visibility. The (iCy3)$_2$ dimer is depicted as light green and dark green line segments attached to the sugar-phosphate backbones (shown in red and blue). The macrostates are color coded: $S_1$ (red), $S_2$ (blue), $S_3$ (green), and $S_4$ (orange). Adapted from Ref. [75].

**III.G DNA breathing of +1, -1 and -2 (iCy3)$_2$ dimer-labeled ss-dsDNA fork constructs and determination of free energy surface parameters**

The combined statistical-dynamical functions shown in Fig. 16 [i.e., the visibility PDF, $P(v)$, two-point TCF, $\bar{C}^{(2)}(\tau)$, and three-point TCF, $\bar{C}^{(3)}(\tau_1, \tau_2)$] serve to constrain a multiparameter optimization of solutions to a four-state kinetic network model, which assumes that the four macrostates at equilibrium interconvert stochastically with forward and backward rate constants. Details of this approach, which have been implemented in previous studies [21, 75, 80, 81], are briefly summarized below.

The time-dependent population of the *i*th macrostate, $p_i(t)$, is modeled using the master equation [85]

$$\dot{p}_i = \sum_{j \neq i=1}^{4} -k_{ij} p_i + k_{ji} p_j \tag{17}$$

Where the $k_{ij(ji)}$ are the forward (backward) rate constants between macrostates $i$ and $j$ [$i, j \in \{1,2,3,4\}$]. A solution to Eq. (17) is obtained for an assumed set of input rate constants, which are subject to completeness and detailed balance conditions. The output parameters from the solution include: (*i*) the equilibrium probabilities, $p_i^{eq}$; and (*ii*) the time-dependent conditional probabilities, $p_{ij}(\tau)$, that the system undergoes transitions from macrostate $i$ to macrostate $j$ during the time interval $\tau$.

Using the output parameters from a solution to Eq. (17), the PDF of the signal visibility is modeled as a sum of four macrostate contributions

$$P(v) = \sum_{i=1}^{4} p_i(v) = \sum_{i=1}^{4} A_i \exp[-(v - \bar{v}_i)^2 / 2\sigma_i^2] \tag{18}$$

where the probability of the *i*th macrostate, $p_i(v)$, is assumed to have a Gaussian form with mean visibility $\bar{v}_i$, standard deviation $\sigma_i$, and amplitude $A_i$. The probability that the *i*th macrostate is present at equilibrium is the integrated area $p_i^{eq} = \int_{-\infty}^{\infty} dv \, p_i(v) = A_i \sigma_i \sqrt{2\pi}$. The two-point and three-point TCFs, respectively, are simulated according to

$$\bar{C}^{(2)}(\tau) = \sum_{i,j=1}^{4} \delta \bar{v}_j p_{ij}(\tau) \delta \bar{v}_i p_i^{eq} \tag{19}$$

and

$$\bar{C}^{(3)}(\tau_1, \tau_2) = \sum_{i,j,k=1}^{4} \delta \bar{v}_k p_{jk}(\tau_2) \delta \bar{v}_j p_{ij}(\tau_1) \delta \bar{v}_i p_i^{eq} \tag{20}$$

where $\delta \bar{v}_i = \bar{v}_i - \langle v \rangle$ is the visibility fluctuation of the *i*th macrostate. The two-point TCF is a time-averaged product of two consecutive observations separated by the interval $\tau$. Similarly, the three-point TCF is the time-averaged product of three consecutive measurements separated by the intervals $\tau_1$ and $\tau_2$, respectively.

For each of the +1, -1 and -2 (iCy3)$_2$ dimer-labeled ss-dsDNA constructs, the statistical functions $P(v)$, $\bar{C}^{(2)}(\tau)$ and $\bar{C}^{(3)}(\tau_1, \tau_2)$ were calculated using Eqs. (18) – (20), respectively, for a given set of input rate constants, $k_{ij}$, mean visibility values, $\bar{v}_i$, amplitudes, $A_i$, and standard deviations, $\sigma_i$. These calculated functions were subsequently compared to the experimentally derived functions. The agreement between simulated and experimental functions was quantified using a nonlinear least squares target function, $\chi^2$, which was minimized by performing an iterative search of the parameter space using a multiparameter optimization algorithm.

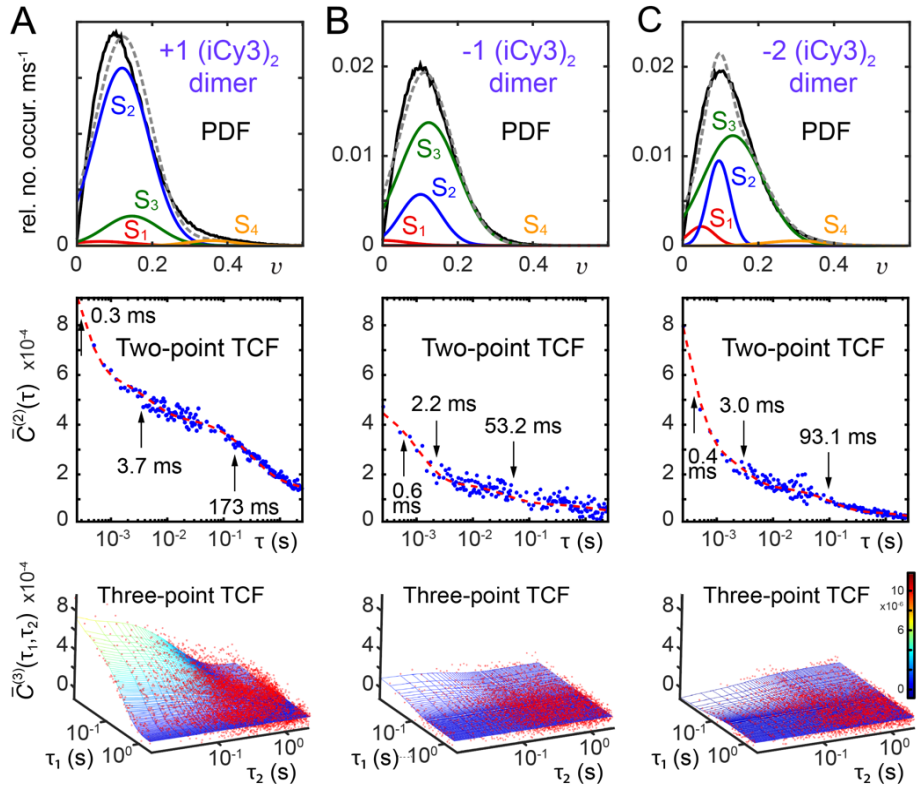

**Figure 19.** Theoretical modeling results of PS-SMF measurements of +1 (**A**), -1 (**B**) and -2 (**C**) (iCy3)$_2$ dimer-labeled ss-dsDNA fork constructs in 100 mM NaCl, 6 mM MgCl$_2$, and 10 mM Tris at pH 8.0. The optimized parameters from a four-state kinetic network model are obtained by simultaneously fitting to the experimental PDF [$P(v)$, black curves, top row], the two-point TCF [$\bar{C}^{(2)}(\tau)$, blue dots, middle row], and the three-point TCF [$\bar{C}^{(3)}(\tau_1, \tau_2)$, red dots, bottom row]. The model PDFs are represented as sums of four Gaussian macrostates (red, blue, green and orange curves, labeled S$_1$ – S$_4$, respectively) with areas equal to the equilibrium probabilities, $p_i^{eq}$, with $i \in \{1,2,3,4\}$. The model two-point TCFs are shown as red curves, and the model three-point TCFs are shown as solid surfaces. Adapted from Ref. [75].

In Fig. 19 is shown a summary of the results of the four-state kinetic network analysis applied to the +1, -1, and -2 (iCy3)$_2$ dimer-labeled ss-dsDNA constructs. In all cases the agreement between simulated and experimental functions is very good. The top row of Fig. 19 shows the visibility PDFs overlaid with the four Gaussian functions that represent the conformational macrostates, $S_1 - S_4$. The normalized areas of the Gaussians are equal to the equilibrium populations of the macrostates. In the second and third rows of Fig. 19, respectively, are shown the two- and three-point TCFs, which characterize the conformational dynamics of the (iCy3)$_2$ dimer-labeled ss-dsDNA constructs as a function of probe labeling position. The two- and three-point TCFs are shown overlaid with their corresponding optimized functions from the network model analysis. It is important to emphasize that determination of the full set of optimization parameters requires the simultaneous fitting of all three statistical-dynamical functions.

As discussed in Sect. III.D, the +1 construct exhibits the slowest overall decay. As the position of the (iCy3)$_2$ dimer probe is varied across the ss-dsDNA junction from +1 to -1, and again from -1 to -2, the relaxation dynamics become faster. This behavior is consistent with the view that the activation barriers that mediate the conformational transitions depend on the relative stabilities of the DNA bases and sugar-phosphate backbones immediately adjacent to the probe. The W-C base pairs within the duplex side of the ss-dsDNA junction are largely stacked, while the stacking interactions between the bases on the ssDNA side of the junction are much weaker due to the lack of complementary base pairing.

In Fig. 20 are shown the optimized kinetic and equilibrium parameters determined for the +1, -1, and -2 (iCy3)$_2$ dimer-labeled ss-dsDNA constructs and the corresponding free energy surfaces. In the left column of Fig. 20 are shown the kinetic network schemes for each of the three constructs. Macrostates $S_1 - S_4$ are shown interconnected by the forward and backward time constants, $t_{ij}$ (= $k_{ij}^{-1}$), along with their equilibrium populations, $p_i^{eq}$. From these parameters are obtained the Boltzman-weighted relative stabilities, $G_{ij} = -k_B T \ln(p_i^{eq}/p_j^{eq})$, and the relative Arrhenius activation energies, $G_{ij}^{\ddagger} = -k_B T \ln(t_{min}/t_{ij})$, where $k_B$ is Boltzmann's constant.

The free energy surfaces shown in the right column of Fig. 20 are useful to visualize the relative stabilities and free energy barriers that separate the four macrostates, which change as the (iCy3)$_2$ dimer position is varied across the ss-dsDNA junctions. In plotting these free energy surfaces, a negative value has been assigned to the visibility of the $S_3$ macrostate corresponding to its left-handed twist, while the remaining macrostates are assigned positive values of the visibility corresponding to right-handed twists. The results indicate that for the +1 construct the $S_2$ macrostate, which is assigned to the W-C conformation, is thermodynamically more stable than the other macrostates, as expected. Furthermore, the kinetic barriers that mediate transitions between $S_2$ and the other macrostates are relatively high, indicating that the W-C conformation at the +1 position is the most mechanically stable. In contrast, for the -1 and -2 constructs the left-handed $S_3$ macrostate is thermodynamically favored and the kinetic barriers that mediate transitions between $S_3$ and the other macrostates at these positions are relatively low. Thus, the left-handed $S_3$ macrostate is thermodynamically favored but mechanically unstable at the -1 and -2 positions.

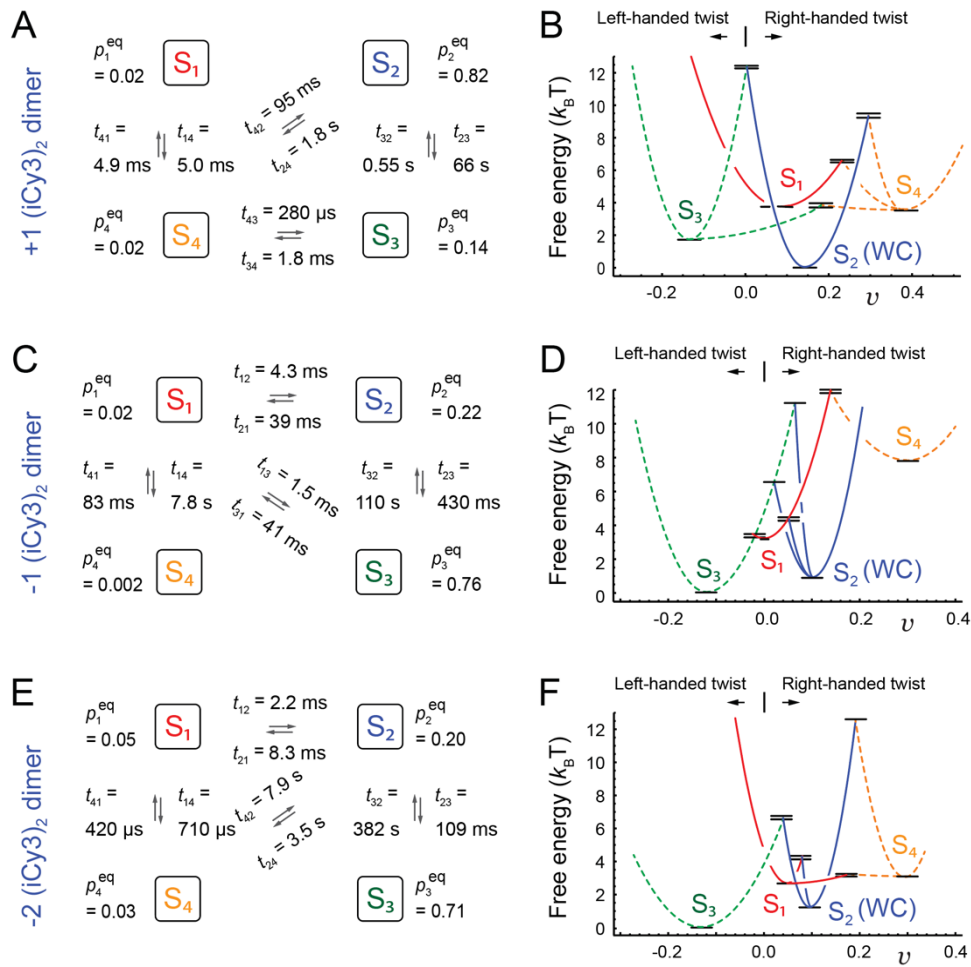

**Figure 20.** Results of kinetic network model analysis of (***A***, ***B***) +1, (***C***, ***D***) -1 and (***E***, ***F***) -2 (iCy3)$_2$ dimer-labeled ss-dsDNA fork constructs. Panels ***A***, ***C***, ***E*** show the kinetic network schemes and equilibrium population, $p_i^{eq}$, and time constant, $t_{ij}$, parameters, which are determined from the analysis of PS-SMF microscopy data. Panels ***B***, ***D***, ***F*** show the corresponding free energy surfaces derived from the optimized equilibrium and kinetic parameters obtained from the analysis. Adapted from Ref. [75].

The above studies show that the bases and sugar-phosphate backbones sensed by the (iCy3)$_2$ dimer probes can adopt four quasi-stable local conformations, whose relative stabilities and transition state barriers depend sensitively on the probe labeling position relative to the ss-dsDNA fork junction. The +1 (iCy3)$_2$ dimer-labeled ss-dsDNA construct exhibits a relatively broad distribution of local base and backbone conformations within the duplex region of the ss-dsDNA junction. The conformational macrostates at this position are dominated by the S$_2$ (W-C) macrostate, which undergoes the slowest dynamics of the three constructs that we studied. This indicates that the S$_2$ macrostate, which is the most thermodynamically stable state within the duplex side of the fork junction, is also the most mechanically stable. In comparison, the -1 and -2 (iCy3)$_2$ dimer-labeled ss-dsDNA constructs exhibit progressively faster dynamics and a distinctly different distribution of conformational macrostates. At these latter positions the left-handed S$_3$ macrostate is most stable, indicating that the thermodynamically favored conformations of bases and backbones sensed by the dimer probe within the ssDNA region of the fork junction are mechanically unstable and distinct from those of the duplex.

While the focus of the current chapter is on the behavior of DNA substrates in the absence of proteins, the structural and kinetic properties of the ss-dsDNA fork junction revealed by such experiments can, in principle, provide mechanistic insights into how replication and repair proteins that assemble at these sites carry out their biological functions. The significance of these studies is that they provide *position-specific* information about the relative stabilities and transition barriers that mediate conformational fluctuations at and near ss-dsDNA fork junctions where such protein-DNA complexes assemble and function.

## IV. Spectroscopic studies of local base stacking conformation using exciton-coupled fluorescent nucleic acid base dimer probes

In previous sections, it was shown that spectroscopic studies performed on exciton-coupled (iCy3)$_2$ dimer-labeled ss-dsDNA constructs can provide detailed, albeit indirect, information about the local conformations and conformational 'breathing' of DNA bases and sugar-phosphate backbones immediately adjacent to the probes. An alternative approach is to use fluorescent base analogues, which can be site-specifically substituted for native bases within a DNA construct. In principle, spectroscopic studies of fluorescent mononucleotide and dinucleotide analogue-substituted DNA constructs can provide direct information about the local conformations of the base analogue dimer probes themselves, and about the local conformations of DNA bases immediately adjacent to the base analogue probe(s) [27, 35, 86, 87].

Base stacking is fundamentally important to the stability of double-stranded DNA. DNA exists in its most stable form as the iconic W-C double-helix in which flanking bases are fully stacked and the sugar-phosphate backbones adopt a right-handed twist [88]. While the W-C structure is the most stable conformation, on local length scales DNA may undergo 'breathing' fluctuations to populate 'thermally excited' conformations in which flanking bases are 'twisted' relative to their neighbors and thus somewhat more or less stacked than in the canonical W-C duplex. In such perturbed conformations, base-pairs are separated, and thus inter-base hydrogen bonds are broken, and/or flanking bases are fully unstacked and un-hydrogen-bonded and thus locally exposed to the solvent environment. Such metastable, non-canonical conformations may aid in the recognition, assembly and function of protein-DNA complexes involved in the core biological processes of DNA replication, transcription and repair [2, 71].

Nucleic acids labeled with fluorescent base analogues such as 6-methyl isoxanthopterin (6-MI, an analogue of guanine) and 2-aminopurine (2-AP, an analogue of adenine) exhibit absorption and CD signals at lower energies than those of the canonical bases (see, e.g., Fig. 2F) and can thus provide information about the local conformations of the probe residues at specific positions within a base sequence [26, 27, 35, 87]. For example, the CD spectrum of a fluorescent dinucleotide-substituted DNA construct, such as the (6-MI)$_2$ dinucleotide substituted DNA construct illustrated in Fig. 2E, can exhibit the spectral signatures of strong exciton-coupling between flanking 6-MI residues [26]. Furthermore, the fluorescence intensity of mononucleotide-substituted DNA constructs can provide additional information about local base stacking interactions and exposure to the aqueous solvent environment [26, 27].

**IV.A Two-photon excitation two-dimensional fluorescence spectroscopy (2PE-2DFS) of the fluorescent guanine analogue 6-MI**

As discussed in Sect. III.B, two-dimensional fluorescence spectroscopy (2DFS) is a fluorescence-detected Fourier transform (FT) optical method that is well suited to measure the electronic-vibrational structure of fluorescent molecular chromophores, such as 6-MI [24, 36, 87]. Like transmission-based multi-dimensional FT optical spectroscopies in the visible and IR regimes [58, 62, 63], 2DFS can reveal information about the optical transition pathways that are operative within a multi-level quantum molecular system. By virtue of its ability to sensitively monitor fluorescence against a dark background, 2DFS can detect relatively weak excited state population signals. Recent advances permit such phase-sensitive fluorescence-detected FT experiments to be carried out under weak signal conditions using single photon counting techniques [67].

Recently, a new experimental approach based on the principles of two-photon excitation (2PE) was introduced to perform 2DFS on the UV-absorbing fluorescent base analogue 6-MI [89]. Like the 2DFS experiments carried out on (iCy3)$_2$ dimer-labeled DNA constructs discussed in Sect. III.B, similar experiments carried out on 6-MI substituted DNA constructs hold promise to provide direct information about local base conformation and conformational distributions. In prior work by Widom *et al.*, 2DFS was used to measure the exciton coupling of (2-AP)$_2$ dinucleotide in solution [87]. Unfortunately, such experiments on UV-absorbing chromophores are limited due to the difficulties of maintaining well-behaved broadband laser pulses in the UV spectral regime and detecting weak fluorescent signals above the background of scattered laser light. These difficulties are circumvented by using a broadband pulsed laser source in the visible regime with wavelength centered at ~675 nm, which is well separated from the wavelength of the fluorescent signal.

Two-photon excitation (2PE) occurs when the photon energy of the source laser is one half of that of the lowest energy transition between ground state, $|g\rangle$, and excited (final) state, $|f\rangle$ (see Fig. 21). While one-photon excitation (1PE) of the $|g\rangle \rightarrow |f\rangle$ transition is not dipole-allowed, 2PE can be achieved through a perturbative sequence of density-matrix elements that include coherences between $|g\rangle$, intermediate 'virtual' states, $|e\rangle$ and $|e'\rangle$, and final states $|f\rangle$ and $|f'\rangle$ [90]. Here, the virtual states are assumed to have energies equal to one half of those of the excited states. Although the optical transitions involving virtual states do not absorb energy from the field, and thus do not lead to fluorescence, the virtual states do participate in coherences, e.g., $|g\rangle\langle e|$, $|e\rangle\langle f|$ or $|e\rangle\langle e'|$. Such 2PE processes have been used to study the electronically excited states of fluorescent DNA base analogues [91].

2PE-2DFS employs similar principles to those of 1PE-2DFS [24, 53-56, 87] in which the sample is illuminated by a repeating sequence of four collinear ultrafast laser pulses, as illustrated in Fig. 21*A*. In these experiments the pulse repetition rate is 250 kHz, the center wavelength is ~675 nm, and the pulse full-width at half-maximum (FWHM) bandwidth is ~30 nm. The photon energy is set to approximately one-half of the lowest energy transition of the 6-MI molecule (with peak absorption wavelength ~340 nm). The pulses are characterized by their center times, $t_i$, phases, $\varphi_i$, inter-pulse delays, $t_{ij} = t_i - t_j$, and relative phases, $\varphi_{ij}$ [$i, j \in \{1,2,3,4\}$]. For a given measurement, the time delays $t_{21}$, $t_{32}$ and $t_{43}$ are held fixed while the relative phases are swept continuously at tens-of-kilohertz frequencies.

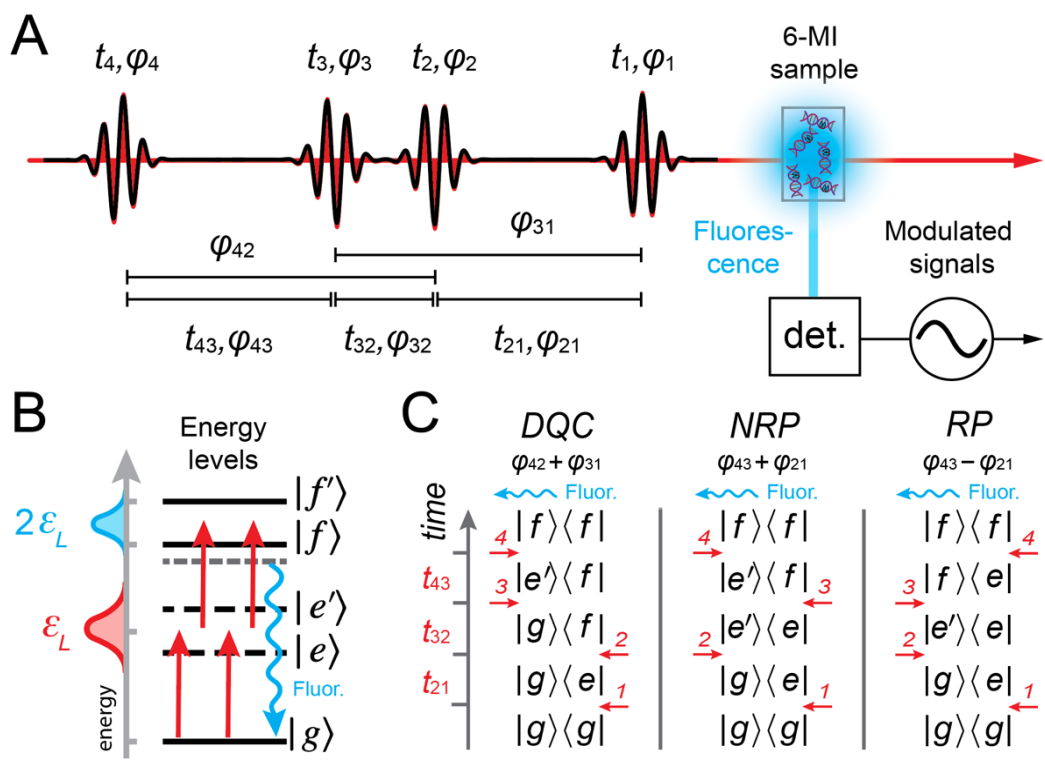

**Figure 21.** (*A*) The 2PE-2DFS method employs a repeating sequence of four collinear ultrafast laser pulses, which excites the 6-MI sample. Fluorescence from the sample is detected using phase sensitive methods. The relevant times and phases are indicated. (*B*) Energy level diagram indicating the laser pulse energy (centered at $\varepsilon_L$) and molecular state energies (centered at $2\varepsilon_L$). Fluorescence is spectrally separated from scattered laser excitation light. (*C*) Double-sided Feynman diagrams for 2PE-2DFS: DQC (double-quantum coherence, $\varphi_{42} + \varphi_{31}$), NRP (non-rephasing, $\varphi_{43} + \varphi_{21}$), and RP (rephasing, $\varphi_{43} - \varphi_{21}$). Adapted from Ref. [89].

The four-pulse sequence induces one-photon transitions between ground state $|g\rangle$, intermediate ('virtual') states, $|e\rangle$ and $|e'\rangle$, and final excited states $|f\rangle$ and $|f'\rangle$, as illustrated in Fig. 21*B* [90]. The ensuing 2PE fluorescence signals are proportional to the populations of the excited states that vary at distinct kilohertz modulation frequencies, which are isolated using phase sensitive detection methods [67]. In Fig. 21*C* are shown double-sided Feynman diagrams, which indicate the three possible pathways of perturbative fourth-order density matrix elements that generate excited state populations. Each of the three Feynman pathways depends on a specific sequence of field-matter interactions that give rise to a unique phase-modulation signature. The 'double-quantum coherence (DQC)' pathway includes the intermediate coherence term $|g\rangle\langle f|$ between ground and excited states during the period $t_{32}$, and has phase signature $\varphi_{42} + \varphi_{31}$. In contrast, the 'non-rephasing' (NRP) and 'rephasing' (RP) pathways, which have phase signatures $\varphi_{43} + \varphi_{21}$ and $\varphi_{43} - \varphi_{21}$, respectively, include the intermediate coherence term $|e'\rangle\langle e|$ between distinct virtual states during the period $t_{32}$. The recorded signals are complex-valued response functions of the inter-pulse delays: $S_{DQC}(t_{21}, t_{32}, t_{43})$, $S_{NRP}(t_{21}, t_{32}, t_{43})$ and $S_{RP}(t_{21}, t_{32}, t_{43})$. Fourier transformation of the response functions with respect to the delay variables yields the frequency- and phase-dependent 2D spectra: $\hat{S}_{DQC}(\omega_{21}, \omega_{32}, \omega_{43})$, $\hat{S}_{NRP}(\omega_{21}, \omega_{32}, \omega_{43})$ and $\hat{S}_{RP}(\omega_{21}, \omega_{32}, \omega_{43})$ [53, 54, 90].

**IV.B Absorption and CD experiments on 6-MI mononucleoside and (6-MI)₂ dinucleotide**

In Figs. 22*A* and 22*B* are shown the molecular structures of 6-MI nucleoside and (6-MI)₂ dinucleotide diphosphate, respectively. The EDTM of the lowest energy ($\pi \rightarrow \pi^*$) $S_1$ transition is indicated as a double-headed blue arrow, which lies within the plane of the 6-MI nucleobase [31] The $S_1$ transition is coupled to at least one effective vibrational mode with energy, $\hbar\omega_{vib}$ = ~400 cm⁻¹. If the two base residues of the (6-MI)₂ dimer are closely spaced – i.e., if they are 'stacked' – the monomer EDTMs can couple electrostatically with Davydov splitting 2$J$ to produce delocalized symmetric (+) and anti-symmetric (–) vibronic excitons, as illustrated in Fig. 22*C*. The precise geometric arrangement of the coupled EDTMs determines the dipole strengths and transition energies of the (6-MI)₂ dimer.

In Figs. 22*D* and 22*E* are shown the experimental absorption and CD spectra, respectively, of the 6-MI mononucleoside (blue) and the (6-MI)₂ dinucleotide (green). The peak absorption energy of the 6-MI nucleoside $S_1$ transition at ~29,300 cm⁻¹ (~341 nm) is significantly lower than that of the second $S_2$ transition at ~34,000 cm⁻¹ (~294 nm). The $S_1$ transition is relatively broad, and its underlying vibronic features are only barely perceptible. Stanley and co-workers resolved vibronic transitions of 6-MI using Stark spectroscopy [30]. The absorption spectrum of the (6-MI)₂ dinucleotide exhibits additional broadening of the $S_1$ line shape in comparison to that of the 6-MI mononucleoside, which is due to the spectral shifts associated with the electrostatic coupling, $J$. Also shown in Fig. 22*D* is the frequency-doubled laser spectrum used in the 2PE-2DFS experiments described below.

The vibronic fine structure is more easily discerned in the CD spectra of the 6-MI mononucleoside and the (6-MI)₂ dinucleotide, which are shown in Fig. 22*E*. In general, the CD spectrum is sensitive to the chiral environment immediately surrounding the EDTM [47, 92, 93]. If the direction of the EDTM depends on the nuclear coordinates of the 6-MI molecule (i.e., a violation of the Condon approximation and the presence of Hertzberg-Teller coupling), the vibronic transitions will give rise to optical activity apparent in the CD spectrum [93-95]. The CD of the 6-MI mononucleoside shown in Fig. 22*E* exhibits a weakly positive band of vibronic features over the range ~26,500 cm⁻¹ to ~32,000 cm⁻¹, corresponding to the $S_1$ transition, and a weakly negative band of vibronic features over the range ~32,000 cm$^{-1}$ to ~33,000 cm$^{-1}$, corresponding to the $S_2$ transition. In contrast, the CD of the (6-MI)$_2$ DNTDP exhibits a strongly positive band (peaked at ~27,000 cm$^{-1}$) and a strongly negative band (peaked at ~30,000 cm$^{-1}$), which symmetrically splits the $S_1$ transition. Such behavior is characteristic of a right-handed Cotton effect that is due to presence of electrostatic coupling between degenerate vibronic states of the component 6-MI monomers [47]. The peak-to-peak splitting is ~2,800 cm$^{-1}$, suggesting an electrostatic coupling $J \lesssim 1,400$ cm$^{-1}$. Like the effect seen for the $S_1$ transition, a right-handed Cotton effect appears to be operative for the $S_2$ transition.

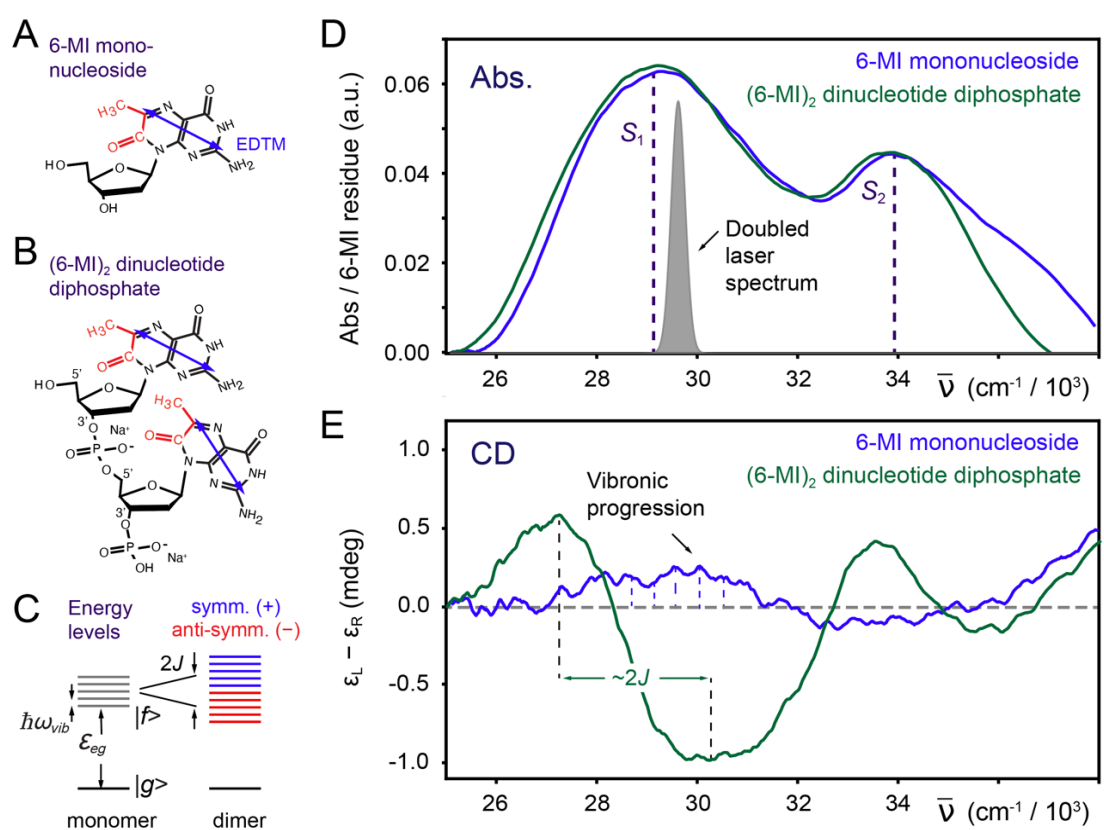

**Figure 22.** (*A*) Molecular structure of the 6-MI nucleoside. Atoms that are additional to those of the natural base guanine are indicated in red. The blue double-headed arrow indicates the direction of the lowest energy electric dipole transition moment (EDTM). (*B*) Structure of the (6-MI)$_2$ dinucleotide diphosphate. (*C*) Energy level diagram of the 6-MI nucleoside and the (6-MI)$_2$ dinucleotide in the presence of electrostatic coupling, $J$, between EDTMs. The monomer electronic transition with energy, $\varepsilon_{eg}$, is coupled to a vibrational mode with energy spacing, $\hbar\omega_{\text{vib}}$. The coupling between monomers introduces symmetric (+, blue) and anti-symmetric (−, red) electronic-vibrational (vibronic) coherent states. (*D*) Absorption spectra of the 6-MI nucleoside (blue) and the (6-MI)$_2$ dinucleotide (green) in aqueous buffer salt solution (10 mM TRIS, 100 mM NaCl, and 6 mM MgCl$_2$). Vertical dashed lines indicate the lowest energy electronic transition ($S_1$) at ~29,300 cm$^{-1}$ and the second lowest transition ($S_2$) at 34,000 cm$^{-1}$ of the 6-MI MNS. The doubled laser spectrum used in the 2PE-2DFS experiments is shown in gray. (*E*) Circular dichroism (CD) spectra of the 6-MI nucleoside (blue) and (6-MI)$_2$ dinucleotide (green). Adapted from Ref. [89].

**IV.C 2PE-2DFS studies of 6-MI nucleoside and (6-MI)$_2$ dinucleotide**

In Fig. 23 are shown the results of 2PE-2DFS experiments carried out on the 6-MI nucleoside. In each panel is plotted the real part of the 2PE-2DFS spectrum for a specific phase condition (from left to right: DQC, NRP and RP) and pulse delay scanning condition, where one of the inter-pulse delays is set to zero (from top to bottom: $t_{32} = 0$, $t_{21} = 0$ and $t_{43} = 0$) while the remaining delay variables are scanned. The 2D spectra are constructed by calculating the Fourier transforms of the measured response functions with respect to the scanned delay variables. The resulting 2D spectra are presented as contour diagrams, in which the horizontal and vertical axes are the frequencies of the scanned delays. In the horizontal and vertical margins of the 2D spectra are shown the linear absorption spectrum of the 6-MI mononucleoside and the frequency-doubled laser spectrum.

The 2D spectra for all nine of the experimental conditions shown in Fig. 23 exhibit peaks and cross-peaks associated with the double-sided Feynman pathways illustrated in Fig. 21*C*. The energy spacings of these features are consistent with the presence of vibronic transitions. For example, all the 2PE-2DFS spectra for $t_{32} = 0$ (top row) exhibit four prominent features with spacing ~400 cm$^{-1}$, which are shifted vertically below the diagonal (along the $\omega_{43}$ axis) by ~200 cm$^{-1}$. These features occur at approximately one half of the transition energies expected of vibronic features associated with the $S_1$ transition, which can be seen in the CD spectrum shown in Fig. 22*E*.

The energies and symmetries of the 2PE-2DFS line shapes [$\hat{S}_{DQC}(\omega_{21}, t_{32} = 0, \omega_{43})$, $\hat{S}_{NRP}(\omega_{21}, t_{32} = 0, \omega_{43})$ and $\hat{S}_{RP}(\omega_{21}, t_{32} = 0, \omega_{43})$] are analogous to those of 1PE-2DFS described in Sect. III.B, except that spectral features that occur at the one-photon transition energies (at the frequencies $\omega_{21}$ and $\omega_{43}$) are associated with coherences between ground state and virtual states (e.g., $|g\rangle\langle e|$ and $|g\rangle\langle e'|$) and between virtual states and final states (e.g., $|e\rangle\langle f'|$, $|e'\rangle\langle f|$). For the 2PE-2DFS spectra with $t_{21} = t_{43} = 0$ (second and third rows of Fig. 23, respectively), prominent spectral features for the DQC phase condition occur at two-photon transition energies (at the frequency $\omega_{32}$) associated with coherences between ground state and final states (e.g., $|g\rangle\langle f'|$, $|g\rangle\langle f|$). This contrasts with the NRP and RP phase conditions in which spectral features occur at relatively low energies at the frequency $\omega_{32}$ associated with coherences between virtual vibronic states (e.g., $|e\rangle\langle e'|$) with vibrational energy spacings.

These results suggest that 2PE-2DFS experiments on the 6-MI MNS can be analyzed using a model Hamiltonian to extract quantitative information about the electronic-vibrational states of the chromophore in its monomeric form.

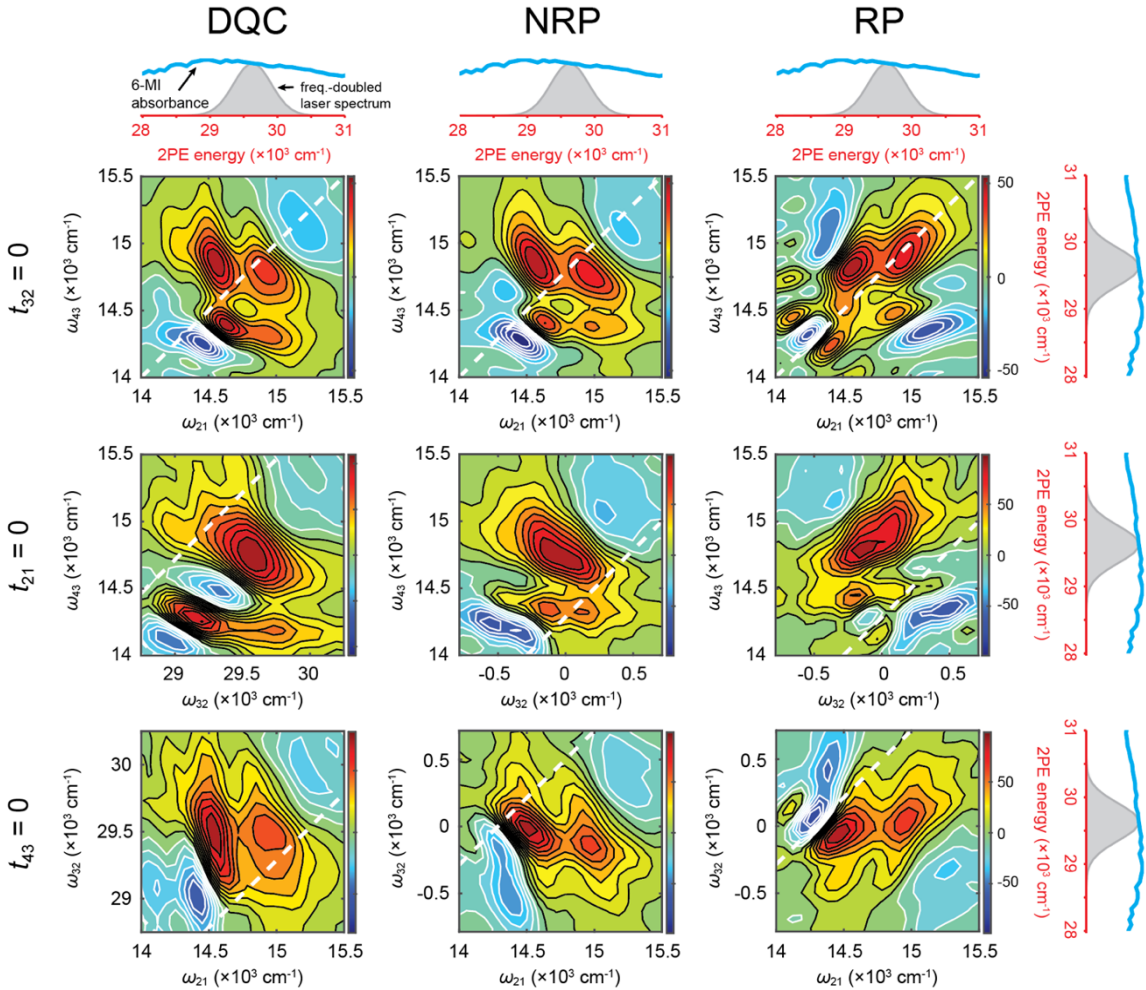

**Figure 23.** 2PE-2DFS experimental results for the 6-MI nucleoside in aqueous buffer salt solution (10 mM TRIS, 100 mM NaCl, and 6 mM MgCl$_2$). The real parts of the experimental spectra are plotted as 2D contour diagrams. Columns indicate a specific signal phase condition (from left to right: DQC, NRP and RP) and rows indicate one of the three inter-pulse delays set to zero (from top-to-bottom: $t_{32} = 0$, $t_{21} = 0$ and $t_{43} = 0$). In the margins are shown the linear absorbance spectrum of the 6-MI MNS (blue) and the spectrum of the laser at twice its energy (gray). Adapted from Ref. [89].

In Fig. 24 are shown the results of 2PE-2DFS experiments carried out on the (6-MI)$_2$ dinucleotide. In general, the 2D spectral features of the (6-MI)$_2$ dinucleotide are narrower, and in some cases split into additional peaks, than the 6-MI nucleoside (compare to Fig. 23). In particular, the 2D spectral features resolved along the $\omega_{32}$ axes for $t_{21} = t_{43} = 0$ (second and third rows) reflect the greatest sensitivity to the presence of exciton coupling. This sensitivity may be related to the absence of obscuring population contributions to the 2PE-2DFS signal during the $t_{32}$ interval (see Fig. 21C). The absence of population contributions to the NRP and RP signals in 2PE-2DFS is therefore a uniquely advantageous situation. In comparison, in 1PE-2DFS the intermediate states are one-photon resonant and population contributions to the fluorescence-detected signal tend to overwhelm any minor contributions associated with coherences. These results for the (6-MI)$_2$ dinucleotide indicate presence of conformation-dependent electrostatic coupling, $J$, between the 6-MI nucleobases within the (6-MI)$_2$ dinucleotide.

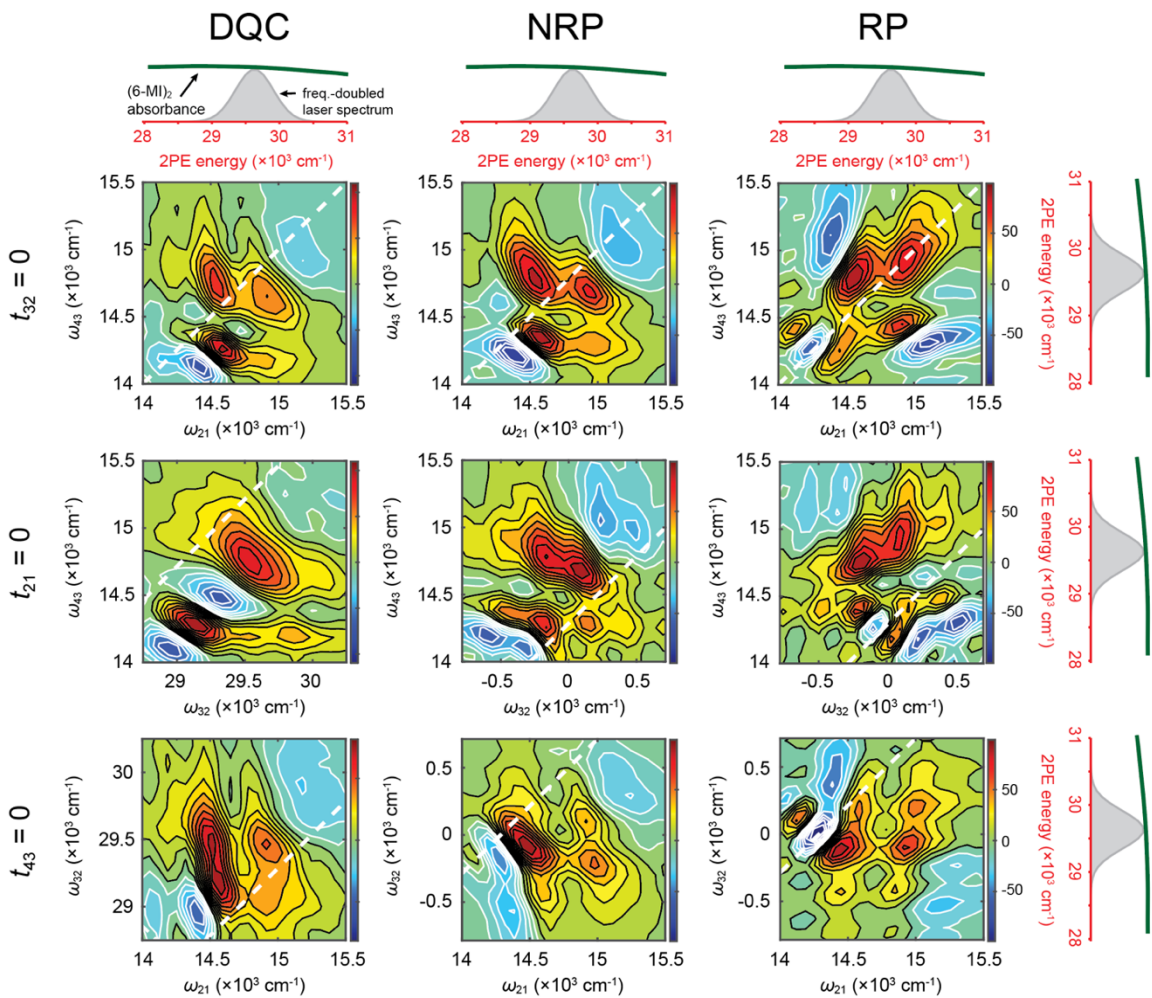

**Figure 24.** 2PE-2DFS experimental results for the (6-MI)$_2$ dinucleotide in aqueous buffer salt solution (10 mM TRIS, 100 mM NaCl, and 6 mM MgCl$_2$). See Fig. 23 caption for further explanation. Adapted from Ref. [89].

The 2PE-2DFS experiments presented in the current section demonstrate that this method can resolve vibronic fine structure of UV absorbing fluorescent nucleic acid base analogues such as 6-MI, which are otherwise difficult to study using standard spectroscopic methods. The sensitivity of the 2PE-2DFS vibronic features to the presence of electrostatic coupling is evident from the differences between the 2D spectra of the 6-MI nucleoside and the (6-MI)$_2$ dinucleotide (compare Fig. 23 to Fig. 24). This sensitivity is likely related to the lack of population contributions to the NRP and RP signals during the $t_{32}$ interval (as illustrated in Fig. 21*C*), which is a unique advantage of the 2PE-2DFS approach. In principle, studies that compare experimental 2PE-2DFS data to theoretical models can determine detailed structural information about local base conformation and conformational disorder in these and related systems. These experiments serve as a proof-of-concept for future investigations of local base conformations of (6-MI)$_2$ dinucleotide-substituted DNA constructs. Such studies hold promise to reveal important information about local base conformations at specific positions in DNA, and their influence on the protein-DNA interactions central to the core biochemical processes of DNA replication, recombination, and repair.

**V. Outlook**

This chapter reviews novel ensemble and single-molecule spectroscopic methods to study the local conformations, conformational disorder and 'breathing fluctuations' of ss-dsDNA fork junctions viewed through the 'lens' of exciton-coupled dimer probes. These methods are highly sensitive to dimer probe conformation due to the coherent coupling between the optical transition densities of the component monomers. Exciton-coupled (iCy3)$_2$ dimer probes, which are site-specifically positioned within the framework of a ss-dsDNA fork construct, provide indirect information about the local DNA bases and sugar-phosphate backbones immediately adjacent to the probes [23-25, 36, 74, 75]. More direct information about the local conformations of the DNA bases can be obtained from DNA constructs in which one or more bases has been site-specifically substituted with a fluorescent base analogue [26, 27, 35, 87, 89].

These systems can be studied using single-molecule fluorescence measurements to directly probe the local conformational fluctuations of the bases and sugar-phosphate backbones of ss-dsDNA constructs over a broad range of time scales, ranging from microseconds to hundreds-of-milliseconds [21, 74, 75, 80, 81]. In addition to providing detailed molecular level information about DNA 'breathing,' and how thermally excited local conformations mediate protein-DNA interactions, such methods can provide insights about the conformational distributions of biologically significant macrostates that exist for a functional protein-DNA complex. In principle, such single-molecule kinetic experiments can provide detailed mechanistic information about the rates of interconversion between stable macrostates, the identities and roles of metastable reactive intermediates, and the connected pathways of protein-DNA complex assembly processes.

Until recently, the relatively unfavorable spectroscopic properties of fluorescent base analogue-substituted DNA constructs (e.g., low fluorescence quantum yields, low absorption cross-sections, inefficient photo-detection at near-UV wavelengths, etc.) has precluded single-molecule optical studies such as those discussed in the current chapter for cyanine probes. Nevertheless, the recent developments in phase-sensitive photon counting detection such as those discussed in the preceding sections [67] suggest that such single-molecule experiments on UV-absorbing molecular dimer probes may soon be feasible.

**Acknowledgements**

The authors are grateful to their current and former laboratory colleagues for many helpful discussions. This work was supported by grants from the National Institutes of Health General Medical Sciences (Grant No. GM-15792 to A.H.M. and P.H.v.H.) and the National Science Foundation RAISE-TAQS Program (Grant No. PHY-1839216 to A.H.M.). S.M. and M.I.S. are supported by the National Science Foundation (Grant No. CHE-2303111). P.H.v.H. is an American Cancer Society Research Professor of Chemistry.